\documentclass[conference,compsoc]{IEEEtran}

\usepackage{nicefrac}
\usepackage{siunitx}
\usepackage{array,framed}
\usepackage{booktabs}
\usepackage{subcaption}
\usepackage{stackengine}
\usepackage{tikz}
\usetikzlibrary{positioning, calc}
\usepackage{
  color,
  float,
  epsfig,
  wrapfig,
  graphics,
  graphicx,
  subcaption
}
\usepackage{textcomp}
\usepackage{amssymb}
\usepackage{tikz}
\usepackage{setspace}
\usepackage{authblk}
\usepackage{pifont}
\usepackage{colortbl}
\usepackage{arydshln}
\usepackage{latexsym,fancyhdr,url}
\usepackage{enumerate}
\usepackage{bbm}
\usepackage{amsmath}
\usepackage{algorithm}
\usepackage{graphics}
\usepackage{xparse} 
\usepackage{xspace}
\usepackage{multirow}
\usepackage{csvsimple}
\usepackage{balance}
\usepackage{xcolor}
\usepackage{cite}
\usepackage{flushend}
\definecolor{takeaways}{HTML}{F5F5F5}
\definecolor{ref}{HTML}{116D6E}
\definecolor{cite}{HTML}{E55807}

\usepackage{
  tikz,
  pgfplots,
  pgfplotstable
}
\usepackage{hyperref}
\hypersetup{                    
  colorlinks,
  linkcolor={green!80!black},
  citecolor={red!70!black},
  urlcolor={blue!70!black}
}
\usepackage{enumitem}

\usetikzlibrary{
  shapes.geometric,
  arrows,
  external,
  pgfplots.groupplots,
  matrix
}

\pgfplotsset{compat=1.9}

\usepackage{mathtools,}
\usepackage{cancel}
\usepackage{dsfont}
\usepackage{float}

\DeclareMathAlphabet{\mathcal}{OMS}{cmsy}{m}{n}

\DeclareGraphicsExtensions{%
    .png,.PNG,%
    .pdf,.PDF,%
    .jpg,.mps,.jpeg,.jbig2,.jb2,.JPG,.JPEG,.JBIG2,.JB2}

\usepackage{xparse}
\newcommand{\bnm}{\begin{newmath}}
\newcommand{\enm}{\end{newmath}}

\newcommand{\bea}{\begin{eqnarray*}}%
\newcommand{\eea}{\end{eqnarray*}}%

\newcommand{\bne}{\begin{newequation}}
\newcommand{\ene}{\end{newequation}}

\newcommand{\bal}{\begin{newalign}}
\newcommand{\eal}{\end{newalign}}

\newenvironment{newalign}{\begin{align}%
\setlength{\abovedisplayskip}{4pt}%
\setlength{\belowdisplayskip}{4pt}%
\setlength{\abovedisplayshortskip}{6pt}%
\setlength{\belowdisplayshortskip}{6pt} }{\end{align}}

\newenvironment{newmath}{\begin{displaymath}%
\setlength{\abovedisplayskip}{4pt}%
\setlength{\belowdisplayskip}{4pt}%
\setlength{\abovedisplayshortskip}{6pt}%
\setlength{\belowdisplayshortskip}{6pt} }{\end{displaymath}}

\newenvironment{newequation}{\begin{equation}%
\setlength{\abovedisplayskip}{4pt}%
\setlength{\belowdisplayskip}{4pt}%
\setlength{\abovedisplayshortskip}{6pt}%
\setlength{\belowdisplayshortskip}{6pt} }{\end{equation}}

\newcounter{ctr}

\newcounter{mytable}
\def\mytable{\begin{centering}\refstepcounter{mytable}}
\def\endmytable{\end{centering}}

\newcounter{myfig}
\def\myfig{\begin{centering}\refstepcounter{myfig}}
\def\endmyfig{\end{centering}}

\newlength{\saveparindent}
\newlength{\saveparskip}
\renewcommand{\eqref}[1]{\mbox{Equation~(\ref{#1})}}

\def \part {part}

\renewcommand{\paragraph}[1]{\vspace*{6pt}\noindent\textbf{#1}\;}

\def \blackslug{\hbox{\hskip 1pt \vrule width 4pt height 8pt
    depth 1.5pt \hskip 1pt}}
\def \qed{\quad\blackslug\lower 8.5pt\null\par}

\newcounter{mynote}[section]

\newcommand\ignore[1]{}

\newcounter{rcnote}[section]

\newcounter{mrnote}[section]

\newcounter{fknote}[section]

\newcounter{anote}[section]

\DeclareMathSymbol{\mlq}{\mathord}{operators}{``}
\DeclareMathSymbol{\mrq}{\mathord}{operators}{`'}

\newcommand{\rhf}[2]{R_{f, \gamma}}

\DeclareDocumentCommand{\edist}{o o}{
  \ensuremath{
    \IfNoValueTF{#1}{{d}}{{\sf d}(#1,#2)}
  }
}

\newcommand{\olrk}[1]{\ifx\nursymbol#1\else\!\!\mskip4.5mu plus 0.5mu\left(\mskip0.5mu plus0.5mu #1\mskip1.5mu plus0.5mu \right)\fi}

\NewDocumentCommand{\indseq}{ O{1} O{r} }{{#1}\ldots {#2}}

\usepackage{xspace,comment}
\usepackage{enumitem} 
\usepackage{amsmath}

\usepackage[utf8]{inputenc} 
\usepackage[T1]{fontenc}    
\usepackage{hyperref}       
\usepackage{url}            
\usepackage{booktabs}       
\usepackage{amsfonts}       
\usepackage{nicefrac}       
\usepackage{microtype}      
\usepackage{xcolor}         
\usepackage{amsmath}
\usepackage{multirow}
\usepackage{arydshln}
\usepackage{makecell}

\usepackage{algorithm}
\usepackage{algpseudocode}

\usepackage{xurl}
\usepackage{subcaption} 
\usepackage{xspace,comment}

\newcommand{\user}{\texttt{Consumer}\xspace}
\newcommand{\oc}{\texttt{Creator}\xspace}
\newcommand{\mg}{\texttt{Modifier}\xspace}
\newcommand{\xclip}{\texttt{X-CLIP}\xspace}
\newcommand{\id}{\texttt{I3D}\xspace}
\newcommand{\mae}{\texttt{MAE}\xspace}

\usepackage[most]{tcolorbox}

\tcbuselibrary{breakable}

\usepackage{threeparttable}
\newcommand{\name}{VGMShield}

\def\thetitle{\name: Mitigating Misuse of Video Generation Models \\
\large An Integrated Approach through Fake Video Detection, Tracing, and Prevention}

\title{\thetitle}

\author[$\dag$]{Yan Pang}
\author[$\S$]{Baicheng Chen}
\author[$\ddag$]{Yang Zhang}
\author[$\dag$]{Tianhao Wang}
\affil[ ]{
        $\dag$University of Virginia \quad $\ddag$CISPA Helmholtz Center for Information Security \protect\\ $\S$ The Chinese University of Hong Kong, Shenzhen
}
\affil[ ]{\{yanpang, tianhao\}@virginia.edu, zhang@cispa.de, baichengchen@link.cuhk.edu.cn}

\begin{document}
\pagestyle{plain}
\maketitle

\begin{abstract}
With the rapid advancement in video generation, people can conveniently use video generation models to create videos tailored to their specific desires. As a result, there are also growing concerns about the potential misuse of video generation for spreading illegal content and misinformation.

In this work, we introduce \name: a set of straightforward but effective mitigations through the lifecycle of fake video generation. We start from \textit{fake video detection}, trying to understand whether there is uniqueness in generated videos and whether we can differentiate them from real videos; then, we investigate the \textit{fake video source tracing} problem, which maps a fake video back to the model that generated it. Towards these, we propose to leverage pre-trained models that focus on {\it spatial-temporal dynamics} as the backbone to identify inconsistencies in videos. In detail, we analyze fake videos from the perspective of the generation process. Based on the observation of attention shifts, motion variations, and frequency fluctuations, we identify common patterns in the generated video. These patterns serve as the foundation for our experiments on fake video detection and source tracing.
Through experiments on seven state-of-the-art open-source models, we demonstrate that current models still cannot reliably reproduce spatial-temporal relationships, and thus, we can accomplish detection and source tracing with over $90\%$ accuracy.

Furthermore, anticipating future generative model improvements, we propose a {\it prevention} method that adds invisible perturbations to the query images to make the generated videos look unreal. 
Together with detection and tracing, our multi-faceted set of solutions can effectively mitigate misuse of video generative models. Our code is available\footnote{\url{https://github.com/py85252876/MMVGM}}.
\end{abstract}


\section{Introduction}

With the success of diffusion models in the field of image generation, video generation has attracted growing interest from the research community. Diffusion-based video generation models have seen substantial development in the past year, with many novel model architectures being introduced. Current public state-of-the-art models, such as Step-Video~\cite{ma2025step} and Hunyuan~\cite{kong2024hunyuanvideo}, are now capable of producing high-resolution and semantically coherent videos. Recently, OpenAI released Sora\footnote{\url{https://openai.com/sora}}, a black-box API-based system that enables users to generate minute-long photorealistic videos.


As video diffusion models rapidly evolve, concerns regarding their misuse cannot be overlooked. 
Malicious individuals have been able to use these models to create and disseminate fake videos online for instigation and malicious propaganda. According to Time Magazine\footnote{\url{https://time.com/7131271/ai-2024-elections/}}, video generation models were used to create misinformation videos that disrupted the electoral process during the 2024 U.S. election.  
In addition to election-related misuse, video generation models can also be exploited to disseminate malicious and illegal content (e.g., child sexual illegal material\footnote{\url{https://www.iwf.org.uk/media/nadlcb1z/iwf-ai-csam-report_update-public-jul24v13.pdf}}).
Therefore, detecting videos generated by such models has become a critical and urgent research problem.
Existing mitigation efforts have focused on deepfakes (generated facial videos)~\cite{gu2021spatiotemporal,khan2021video,ciftci2020fakecatcher, li2021deepfake} as well as other modalities, like image~\cite{dang2020detection,ni2022core, bonettini2020video, hu2021dynamic, gu2021spatiotemporal,sha2023defake,ha2024organic} and text~\cite{krishna2023paraphrasing,mitchell2023detectgpt}. However, the misuse of samples generated by general-purpose video generation models beyond facial deepfakes remains largely unexplored in current literature. We will talk about these in more detail in~\autoref{sec:related_work}. 

In this work, we propose \name, a set of straightforward but effective mitigation strategies through the lifecycle of fake video generation. Our approach analyzes the video generation process, including attention shift, motion flows, and frequency fluctuations. We observe that generated videos often suffer from several quality issues, including attention shifts at both the denoising step and frame level, as well as motion inconsistency and irregular frequency patterns. These issues manifest as unnatural motion transitions and a lack of coherent high-frequency content, which are indicative of temporal instability and low perceptual fidelity. We collectively refer to these features in generated videos as spatial-temporal dynamics. We then leverage pre-trained video recognition models to detect spatial-temporal dynamics in generated content. We first delineate three roles based on the life-cycle of generated content: \oc, \mg, and \user. 
Initially, there (optionally) exists original content created by \oc, mostly for benign purposes like sharing. The malicious \mg then takes the generative model to create fake content (in our context, videos). Finally, \user reads those contents. We have a more detailed discussion in \autoref{fig:demo} in \autoref{sec:background}. 

For \user, we design \textit{detection} to empower them in distinguishing fake videos. We consider three detection models that use different pre-trained video recognition models to extract spatial-temporal features. These pre-trained models serve as the backbone, linked to fully connected layers for detection, upon evaluating these detection models in four detection scenarios that mirror real-world conditions. We categorize scenarios based on the background knowledge of the model and data, as detailed in~\autoref{sec:detection_meth_detect}. Notably, \mae-based detection model consistently outperforms the other detection models.


Next, we consider the {\it source tracing} problem that identifies which model the generated video comes from. The intuition is that different models exhibit different features when generating videos. Tracing can also potentially help with \textit{the regulation of generative models} (by identifying which models are being misused). Similar to building our detection models, tracing models are also based on pre-trained video recognition models as backbones. \mae-based models show effectiveness in tracing, can achieve $97\%$ accuracy in \textit{data-aware} setting. Even in the more realistic setting, it can still achieve $90\%$ accuracy.



To investigate why our detection and source tracing models are effective and the reasons behind performance differences across different backbone models, we then employ the visualization techniques (i.e., Grad-CAM~\cite{Selvaraju_2019} and attention map visualization~\cite{chefer2021transformer}) for a detailed analysis. Both techniques are widely used machine learning explainability methods that help understand why models make specific decisions on inputs. It highlights regions of the input that receive more attention during the model's execution (more details in~\autoref{sec:detailed_analysis}). By applying Grad-CAM and visualizing the attention map to several representative samples, we observe distinct traits of the \mae-based detection model~\cite{tong2022videomae}. It shows versatility in detection capabilities and heightened sensitivity to temporal distortions.


Finally, for \oc, we introduce \textit{misuse prevention} to disrupt generation, thereby safeguarding the integrity of content originated by \oc. The basic idea is to add perturbations to the image, making it unsuitable as a query input for VGMs. Video and image generation models behave differently due to processing differences. The motion prediction term needs to be considered in our work. We designed two defense strategies within our setting, demonstrating robust defensive capabilities in our experiments. Our comprehensive pipeline is evaluated on two publicly available high-quality video datasets. It encompasses seven open-source and two commercial video generation models, covering eleven distinct generation tasks.

 
\paragraph{Contributions.}
The contributions of our work are:

\begin{itemize}[leftmargin=*]

    \item 
    Our defense pipeline is specifically designed for samples generated by general video generation models, comprising three key components: \textit{detection}, \textit{source tracing}, and \textit{prevention}. 
    The \textit{detection} component comprehensively considers four real-world scenarios and is designed with four distinct variants. The \textit{source tracing} model can trace the origin of a video based on subtle differences in the content. 
    Meanwhile, \textit{prevention} offers two different defense methods, both providing effective protection against various video generation models.
    
    \item Our work systematically evaluates the effectiveness of the proposed methods, incorporating two open-source datasets, seven open-source models, and eleven (including both text-to-video and image-to-video) generative tasks. 
    Furthermore, we designed an adaptive attack to showcase the robustness of our defense strategies.
    
    \item Before the experiments, we conducted a preliminary analysis on generated videos using PCA and frequency-domain analysis. We observed attention shifts, motion variations, and spectral fluctuations that commonly appear in generated content. Following the experimental results, we performed a qualitative analysis on representative samples. By employing Grad-CAM and attention map visualization, we identified key patterns that influenced the decisions of both the \textit{detection} and \textit{source tracing} models.

\end{itemize}


\section{Background}\label{sec:background}
\subsection{Denoising Diffusion Generation Models} \label{sec:background_diffusion}
Diffusion models~\cite{ho2020denoising} encompass two primary processes: the forward diffusion process and the reverse denoising process, which progressively removes noise from an image, ultimately generating the final output.

The forward process can be conceptualized as a Markov chain. Starting with the input image $x_0$, the noisy image at time $t$, denoted as $x_t$, is dependent solely on the noisy output from the previous moment, $x_{t-1}$:
\begin{equation}
    x_t = \sqrt{\alpha_t}x_{t-1} + \sqrt{1-\alpha_t} \epsilon; \quad\epsilon \sim \mathcal{N}(0,1)\label{x_t}
\end{equation}
where $\alpha_t$ is a pre-defined noise schedule.
Subsequently, employing the reparameterization trick enables the direct derivation of the noised image at time $t$ from the original image $x_0$, which can be expressed as follows:
\begin{equation}
    x_t = \sqrt{\bar{\alpha}_t} x_{0} + \sqrt{1 - \bar{\alpha}_t} \epsilon_t;\quad\epsilon_t \sim \mathcal{N}(0,1) \label{x_0}
\end{equation}
%
%
In the denoising process, a neural network (e.g., UNet) $\epsilon_{\theta}$ is trained to predict $\epsilon_t$ given the input $x_t$ and time step $t$, thereby achieving a reduction in noise level to obtain $x_{t-1}$. Only the denoising process is needed in the inference process. The diffusion process is used to get $x_{t}$ during training the $\epsilon_{\theta}(x_t,t)$:
%
%
%
\begin{equation}
    L_t(\theta) = \mathbb{E}_{x_0,\epsilon_t} \left[ \lVert \epsilon_t - \epsilon_{\theta}(\sqrt{\bar{\alpha}_t} x_0 + \sqrt{1 - \bar{\alpha}_t} \epsilon_t, t) \rVert^2_2\right]\;
    \label{eq:loss}
\end{equation} 

We also provide more details about diffusion models at~\hyperref[appendix:diffusion]{Appendix~\ref*{appendix:diffusion}}.

\paragraph{Video generation using Diffusion Models.}
Videos are essentially sequences of images. Current video generation models predominantly adopt the architecture of diffusion models with temporal layers for video synthesis~\cite{zhang2023i2vgenxl, zhang2023show1, ho2022video, singer2022makeavideo, ho2022imagen,zhou2023magicvideo, he2023latent, chen2024videocrafter2overcomingdatalimitations, blattmann2023align, wang2023videocomposer,wang2023lavie,blattmann2023stable, esser2023structure,chen2023seine, Mullan-Hotshot-XL-2023}. 
Early video diffusion models of this kind inherit their {\it spatial} domain understanding from diffusion models, and integrate a {\it temporal} convolution layer into each UNet block, to produce videos. In contrast, recent state-of-the-art video generation models~\cite{ma2025step, kong2024hunyuanvideo} use DiT architecture, which replaces the traditional UNet with a pure Transformer architecture, enabling end-to-end spatiotemporal modeling and improved temporal consistency.  

\autoref{tab:video_generative_model} summarizes nine generative tasks across seven state-of-the-art open-source video generation models. These models accept prompts or images through an encoder (e.g., \texttt{CLIP}~\cite{radford2021learning}) as conditional inputs to guide the generation of videos with frame numbers ranging from $16$ to $96$. The generated videos' duration is from $2$ to $4$ seconds. 

Note that there are also video {\it modification} models that take videos as input. These models can modify the object or motion depicted in the original footage~\cite{wu2023tuneavideo,liu2023videop2p,zhao2023motiondirector,shin2023editavideo, karim2023save, molad2023dreamix, ceylan2023pix2video, zhang2023controlvideo, zhao2023controlvideo}. For instance, video editing models are capable of transforming the content from `A man is playing basketball' to `A panda is playing basketball' throughout the video. This work focuses on generative models that take images and/or text as input. These {\it modification} models are widely applied in deepfake tasks, where the goal is to replace faces in videos to create realistic forgeries.

\begin{table}[t]
    \centering
    \caption{Representative open-source generative tasks, detailing for task category, video resolution, and frame rate. `I' refers to Image, and `T' denotes Text.}
    \vspace{0.1cm}
    \label{tab:video_generative_model}
    \Huge
    \resizebox{0.47\textwidth}{!}{
    \begin{tabular}{ccccc}
    \toprule[2.75pt]
        Model & Open Sourced & Input  & Video Resolution & \# Frames \\ 
    \midrule[2pt]
        \multirow{2}{*}{\begin{tabular}[c]{@{}c@{}}VGen~\cite{zhang2023i2vgenxl} \end{tabular}} & \ding{51} & T  & $448\times256$ &  $16$  \\
             & \ding{51}  & I + T  & $1280\times704$ &  $16$   \\
    \hdashline[0.8pt/5pt]
        Lavie~\cite{wang2023lavie} & \ding{51} & T & $512\times320$  & $16$ \\
    \hdashline[0.8pt/5pt]
        Seine~\cite{chen2023seine} & \ding{51}  & I  & $560\times240$  &  $16$\\
    \hdashline[0.8pt/5pt]
        Stable Video Diffusion~\cite{blattmann2023stable}& \ding{51}   &  I & $1024\times576$ &   $25$  \\
    \hdashline[0.8pt/5pt]
    Hunyuan Video~\cite{kong2024hunyuanvideo} & \ding{51} & T & $544\times960$ & $130$ \\ \hdashline[0.8pt/5pt]
    Step-Video-T2V~\cite{ma2025step} & \ding{51} & T & $544\times992$ & $17$ \\ \hdashline[0.8pt/5pt] 
        \multirow{2}{*}{\begin{tabular}[c]{@{}c@{}}VideoCrafter~\cite{chen2024videocrafter2overcomingdatalimitations} \end{tabular}}& \ding{51}  & I + T  & $512\times320$ &  $16$\\
          & \ding{51}  &  T  & $1024\times576$ & $16$ \\
          \hdashline[0.8pt/5pt]
          Gen-2\footnotemark{} & \ding{55} & I + T & $768 \times 448$  & $96$ \\
          Pika Lab\footnotemark{} & \ding{55} & I + T  & $1024 \times 576$ & $72$ \\
        \bottomrule[2pt]
    \end{tabular}}
\end{table}

\addtocounter{footnote}{-2} 
\stepcounter{footnote}\footnotetext{\url{https://research.runwayml.com/gen2}}
\stepcounter{footnote}\footnotetext{\url{https://pika.art/}}

\subsection{Problem Statement}\label{Threat_Model}

\begin{figure}[t]
    \centering
    \includegraphics[width = 0.47\textwidth]{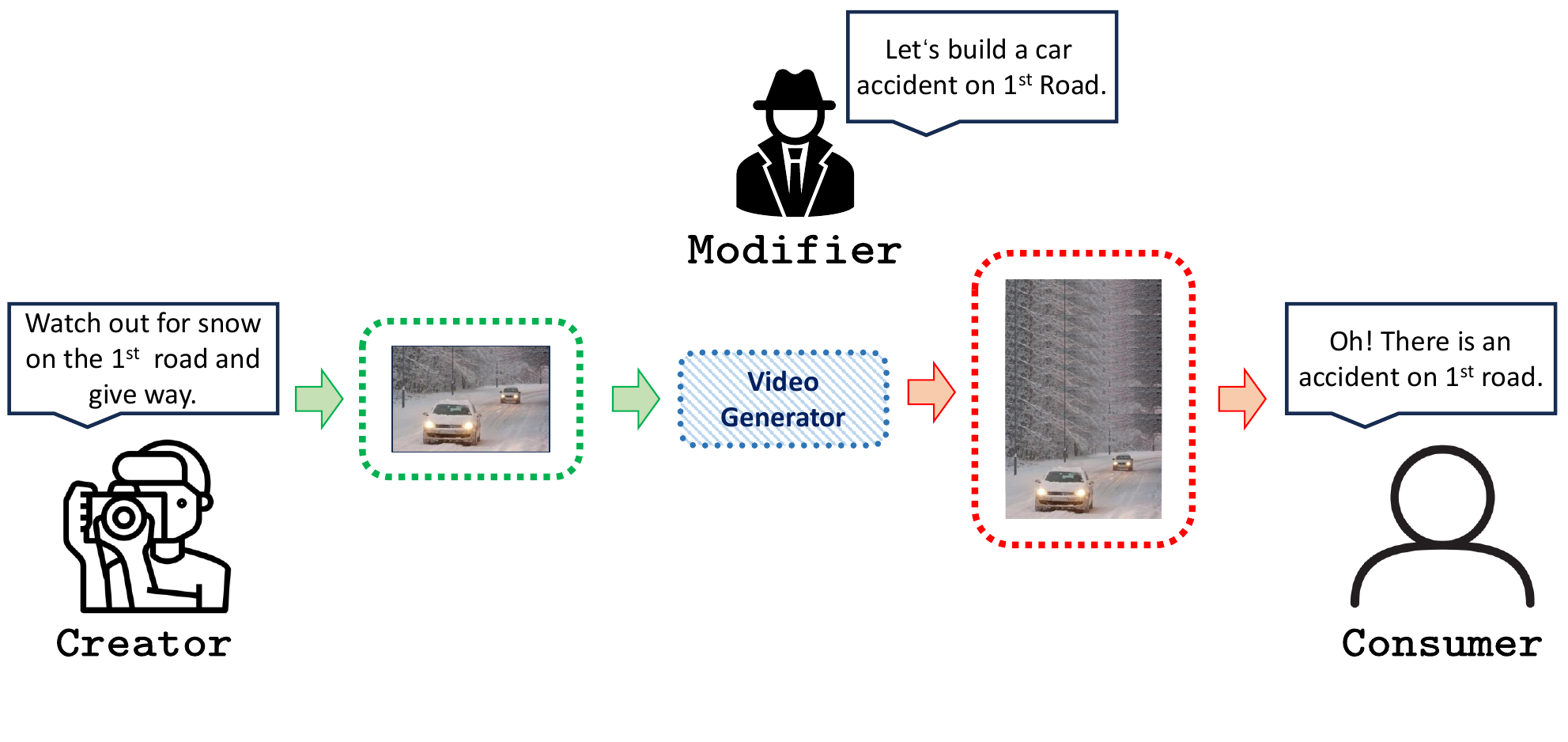}
    \caption{We assume there are three parties: \user, \oc, and \mg. In a typical scenario, \oc creates images, e.g., a road with snow to notify people to take care, and publishes them; \mg takes that content and creates videos for malicious purposes, e.g., a video of a car accident; when \user sees the malicious videos, they may be scared.}
    \label{fig:demo}
\end{figure}

We start by modeling parties in real-world scenarios into three distinct entities: \oc, \mg, and \user, following the life-cycle of information/content generation and consumption. \autoref{fig:demo} provides a demonstration of the roles of these three entities.
%
For example, photographers or journalists can be the \oc. They upload information for the \user. However, due to the presence of \mg, a portion of the images they upload may be maliciously used to generate fake videos to mislead \user (e.g., they could have topical controversy and sway public opinion). 



To address the safety concerns posed by video generation models, we propose a comprehensive defense framework comprising three distinct approaches:

\begin{itemize}[leftmargin=*]
    \item \textit{Detection}: Detection informs a \user about the authenticity of the videos they are viewing, discerning whether they are AI-generated. 
    \item \textit{Source tracing}: Tracing aims to inform a \user about which specific video generation model produced a given video after it has been identified as a fake video.
    Technically, detection and tracing are both classification tasks; we will discuss them together. 
    \item \textit{Misuse prevention}: For defense against image-to-video generative tasks, we introduce our method that adds perturbations across both spatial and temporal dimensions to safeguard image assets, thereby preventing video generation models from successfully synthesizing videos from these input images.
\end{itemize}

\subsection{Deepfake Video Detection} 
One closely related area of research is deepfake detection. The early definition of deepfake videos refers to those manipulated by humans, using not only deep learning models but also graphics-based methods~\cite{guera2018deepfake}. Guera et al. observed the development of generative models~\cite{goodfellow2014generative} and face-based video manipulation techniques~\cite{thies2016face2face}. These advances led to the creation of face-swap videos, which can deceive people. To address this issue, they designed the first deepfake video detection model, which used CNNs to extract frame-level features and RNNs for classification. Following their work, many similar approaches have been proposed~\cite{gu2021spatiotemporal,khan2021video,ciftci2020fakecatcher}. 

Today, many advanced video generation models~\cite{zhang2023show1,zhang2023i2vgenxl,chen2024videocrafter2overcomingdatalimitations, blattmann2023stable,chen2023seine,wang2023lavie} have been developed. 
The fake videos produced by these new models are significantly different from those targeted by earlier works~\cite{gu2021spatiotemporal,khan2021video,ciftci2020fakecatcher, li2021deepfake,guera2018deepfake}. Our study aims to address this gap by focusing on the privacy issues posed by models capable of generating more realistic and diverse videos.

\begin{table}[t]
    \centering
    \caption{Summary of different scenarios for detection and tracing. `Data' indicates the distribution of the data source (e.g., the fake videos are generated from images from a specific movie ), and `Model' indicates the generating model. \ding{51}: Known, \;\ding{55}: Unknown.
    }
    \label{tab:scenario_conclude}
    \resizebox{0.47\textwidth}{!}{
    \begin{tabular}{cccc}
    \toprule
         Task & Setting
         & Data & Model  \\
         \midrule
         \multirow{4}{*}{\begin{tabular}[c]{@{}c@{}} Detection {} \end{tabular}}  & Targeted detection  & \ding{51} & \ding{51}  \\
         & D-blind & \ding{55} & \ding{51} \\
         & M-blind  & \ding{51} & \ding{55} \\
         & Open detection & \ding{55} & \ding{55} \\
         \midrule
        \multirow{2}{*}{\begin{tabular}[c]{@{}c@{}} Source Tracing {} \end{tabular}} & Data-aware & \ding{51}  & \ding{55}  \\
         & Data-agnostic & \ding{55} & \ding{55} \\
    \bottomrule
    \end{tabular}}
    \vspace{0.2cm}
\end{table}

\section{Fake Video Detection and Tracing} \label{sec:fakevideodetection}
In the realm of image generation models, it has been observed that generated images show noticeable differences from real ones in the semantic distribution~\cite{sha2023defake}. Compared to images, we consider generated videos to be more complex and information-rich. Therefore, we hypothesize that differences within generated videos, as well as between generated and real ones, are more detectable. We posit that this observation may generalize to video generation: \textit{videos produced by generative models tend to exhibit unique, model-specific characteristics across spatiotemporal dimensions}. We aim to leverage these traits for video detection and source tracing. Firstly, we analyze and categorize different scenarios.

\subsection{Threat Model} 


\subsubsection{Detection}\label{sec:detection_intuition}
We categorize the task of detection based on the availability of two types of background knowledge: \ding{182} the origin of the input data, and \ding{183} from which model the target video is generated. These two types of information are not always reasonable assumptions in real-world scenarios; we consider them more for a comprehensive understanding of the technique. 
The four settings are summarized in~\autoref{tab:scenario_conclude}.

\paragraph{Targeted Detection.} In this scenario, detectors have knowledge of the potential models that could have generated a given video (if it is an AI-generated video). Additionally, the data distribution (input prompt/image) used to generate this video is also informed. This scenario is highly idealized. 

\paragraph{D-blind.}
In this setting, detectors may know which model is used but lack information about the data (image and/or text) distribution that is used to generate fake videos. For example, detectors might know that \texttt{Hunyuan}~\cite{kong2024hunyuanvideo} is used to generate fake videos on specific topics, because it is the current state-of-the-art open-sourced model. 
We simulate the D-blind setting by training the detection method on one real/fake video dataset but testing it on another dataset.



\paragraph{M-blind.} Similarly, {M-blind} indicates situations where the generation model is unknown, but the source of the data (distribution) is known.

\paragraph{Open Detection.}
Lastly, {open detection} considers the most challenging and perhaps most realistic scenario, where both the data source and the model are unknown.



\begin{figure}[t]
    \centering
    \includegraphics[width=3.4in]{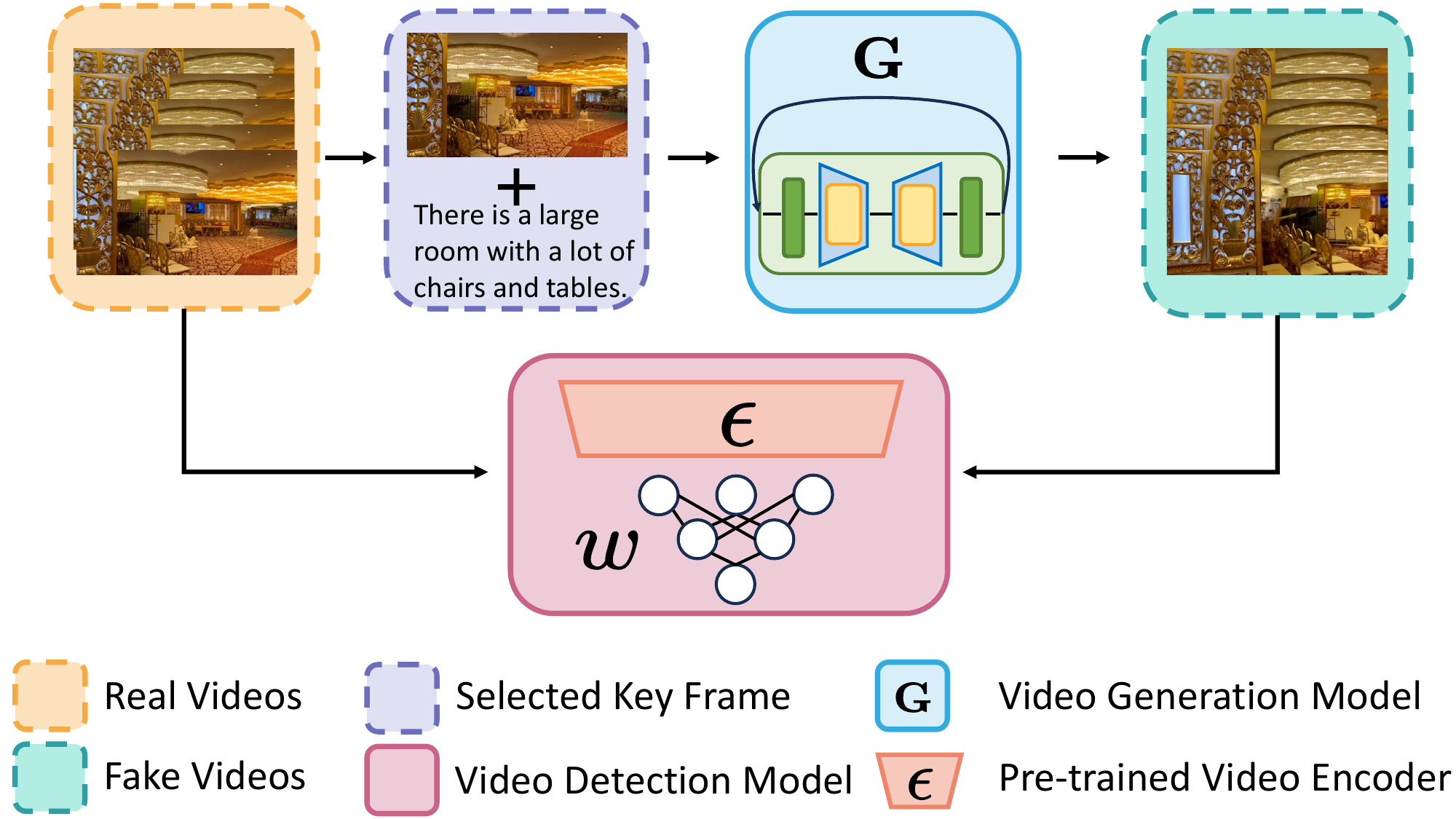}
    \caption{Experimental pipeline for \textit{detection}. To make sure real and fake videos follow similar distribution, we generate fake videos using the first frame of a real video (and optionally the associated caption). The classification model is composed of a video recognition model as the backbone and a fully connected model,
    trained on real/fake videos.}
    \label{fig:fake_video_detection}
\end{figure}
\vspace{-5mm}
\subsubsection{Source Tracing}

Tracing leverages the characteristics of fake videos to locate the origin of the video generation model. In the source tracing task, the test sample is already identified as a fake video. Since the goal is to trace its origin and the identity of the generative model is not available by default, the task naturally operates under the M-blind setting, resulting in two key scenarios. To differentiate from detection tasks, we call them {\it data-aware} and {\it data-agnostic} settings (bottom two rows in~\autoref{tab:scenario_conclude}).

\subsection{Method}

We have reformulated detecting fake videos and tracing fake video generation models as a classification task. We postulate that {\it fake videos exhibit spatial anomalies and manifest temporal inconsistencies and anomalies}. Hence, we adopt pre-trained video recognition models with capabilities to understand spatial and temporal dynamics to serve as the backbone for our detection and source tracing models. 

Denote the pre-trained video recognition model as $\epsilon$ ($\epsilon$ can be \id, \xclip, or \mae), we straightforwardly connect them with trainable, fully connected layers $w$, and obtain the final detection model $f=w\cdot\epsilon$ (to denote $f(x)=w(\epsilon(x))$.
During the training phase, we modify both $\epsilon$ and $w$.

\subsubsection{Detection} \label{sec:detection_meth_detect}
Given the base model $f$, the detailed constructions to handle different scenarios (listed in~\autoref{tab:scenario_conclude}) differ only in how to train the model. There are two generic principles that probably apply to all classification problems: 
\begin{itemize}[leftmargin=*]
    \item {\it The training set should be as diverse as possible. } This applies to the open detection setting, where we curate as many real and fake videos as possible (The number of real and synthesized videos is equal).
    \item 
    {\it When the task is more specific, the training set should be narrowed to match the task.} This applies to the settings when the detector knows about the model being used and/or the data distribution to generate the videos. In those settings, the training set only includes videos generated from the same model and/or the same distribution.
\end{itemize}

More concretely, as will be detailed later in~\autoref{sec:detection_eval}, we have $G=\{G_0,\ldots,G_8\}$ of $9$ tasks and $D=\{D_0, D_1\}$ of $2$ video datasets. For each real video in the video dataset, we generate its corresponding fake videos. Specifically, we use the images (first frame) and (if applicable) captions to query each video generation model to produce fake videos. This is to minimize the distance between real and fake videos (to minimize the detector's reliance on other features, e.g., real videos are always about animals while fake ones are always about cars). The paradigm is shown in~\autoref{fig:fake_video_detection}.

Once the training set is constructed, we can train $f$ for different scenarios following the above-mentioned principles. To simulate the case where the data distribution is unknown, we train the detector $f$ on real and fake videos from one dataset $D_b$ ($b \in {0,1}$), and test it on the other dataset $D_{\bar b}$. To simulate the case where the generation model is unknown, we train $f$ using real videos and fake videos generated by a specific model $G_t$, and test it on fake videos generated by the remaining models in the set $G \setminus G_t$. For open-set detection, we train $f$ using data from $D_b$ and a subset of $9$ tasks $G_s \subset G$, and test on fake videos generated by models in $G \setminus G_s$ using query data from $D_{\bar b}$. 

\begin{figure}[t]
    \centering
    \includegraphics[width=\linewidth]{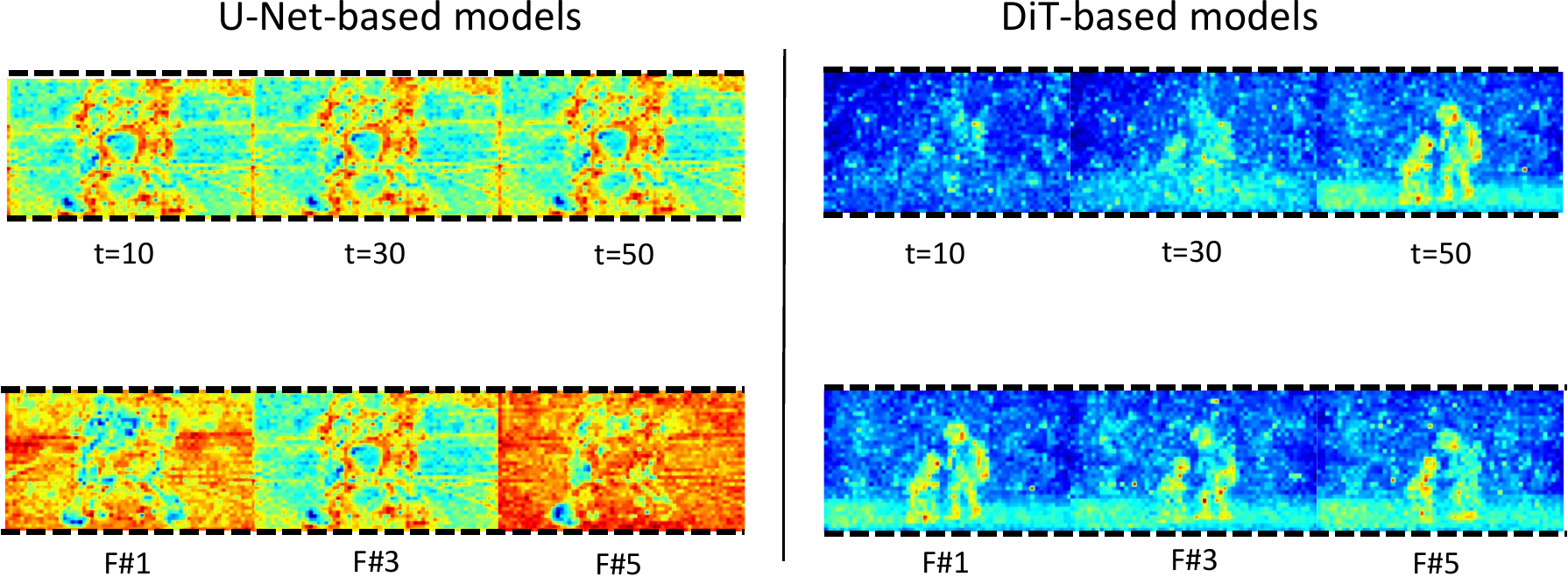}
    \caption{Presenting temporal (across frames) and inference-level (across denoising steps) attention shifts in video generation models. Left: VideoCrafter; Right: Hunyuan.}
    \label{fig:model_generation_vis}
\end{figure}

\subsubsection{Source Tracing} \label{sec:source_tracing_methodology}


Technically, \textit{source tracing} is very similar to \textit{detection}, as they are both classification models. However, there are some differences.

First, tracing assumes the data (video) is always fake, and $f$ becomes a {\it multi-label} classification model. Second, because here the task is to guess which one of the $9$ models the fake video comes from, although the model is set to unknown, the training of $f$ always involves fake videos generated from all $9$ models. This is different from the M-blind setting in detection, where the generative model used in testing is not used during the training process.

\begin{figure}
    \centering
    \includegraphics[width=0.40\textwidth]{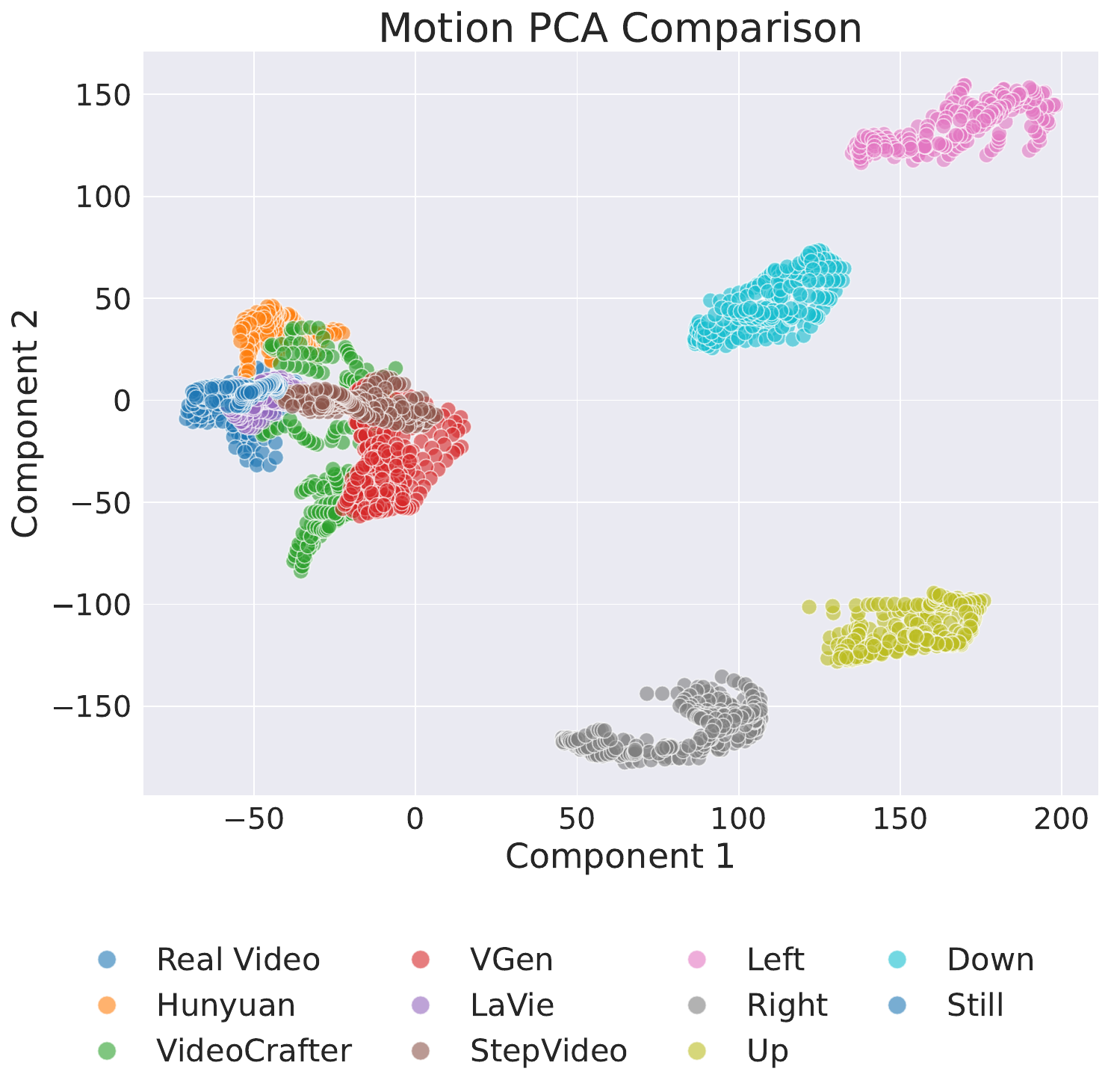}
    \caption{Comparison of motion consistency and prompt alignment across video generation models. StepVideo~\cite{ma2025step} and Hunyuan~\cite{kong2024hunyuanvideo} perform well, while early models like VGen~\cite{zhang2023i2vgenxl} and VideoCrafter~\cite{chen2024videocrafter2overcomingdatalimitations} struggle.}
    \label{fig:motion}
    \vspace{-0.5cm}
\end{figure}

\subsection{Preliminary Analysis} \label{sec:preliminary_analysis}

In this section, we further explore several spatial-temporal features that can help distinguish generated videos from real ones and from other generated videos.

\subsubsection{Attention Shift} The first feature we focus on is attention shift, which typically refers to a behavior observed in diffusion-based models during generation. Since the diffusion model performs generation through a step-by-step denoising process, it initially concentrates on forming the overall structure and coarse textures. As the denoising timestep decreases, the model gradually adds finer details.

This generation feature results in a clear temporal progression in the emergence of high-frequency textures, which appear incrementally over time. We refer to this as temporal dynamic change. Because generative models struggle to maintain temporal coherence in high-frequency features like edges and textures, the high-frequency energy in generated videos is more spatially and temporally dispersed than in real ones. We analyze this phenomenon in detail in~\autoref{freq_analsis}.

Furthermore, we compare attention distributions across different frames at the same timestep. Interestingly, we find that the model tends to shift its attention to different spatial regions even between adjacent frames. This frame-level variation in attention can introduce temporal inconsistencies, such as unnatural flickering or unstable object details, which are typical artifacts of the generation process.

This pattern aligns with what we observe in the generation process of Hunyuan~\cite{kong2024hunyuanvideo} and VideoCrafter~\cite{chen2024videocrafter2overcomingdatalimitations}. Attention maps in early diffusion stages highlight semantically important regions, guiding the spatial layout of the frame. In later stages, the model amplifies high-frequency signals, such as texture and edge details, which contribute to the final visual appearance but may also exacerbate temporal inconsistencies. 

\begin{figure*}[!h]
    \centering
    \includegraphics[width=0.98\textwidth]{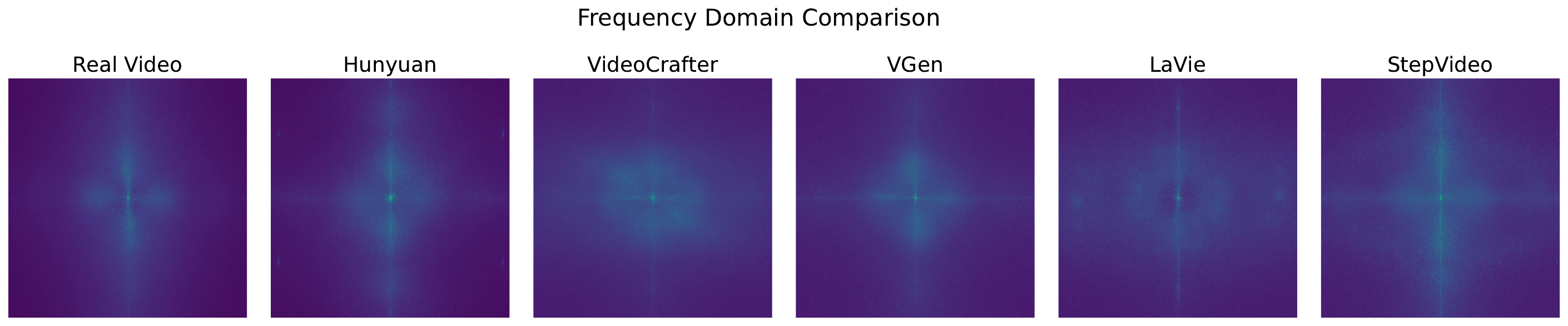}
    \caption{Comparison of high-frequency signals and spectral patterns in real vs. generated videos, indicating weaker texture fidelity and temporal consistency in generated videos.}
    \label{fig:frequency}
\end{figure*}

\subsubsection{Motion Variation} In this part, we conduct a motion analysis of the videos generated by different models, following the pipeline of Xiao et al.~\cite{xiao2024videodiffusionmodelstrainingfree}. Our goals are twofold: (i) to assess the consistency of motion across generated videos and (ii) to evaluate each model’s ability to follow motion‐related instructions specified in the prompts.

As shown in~\autoref{fig:motion}, all models generally follow prompts that explicitly specify object motion. For example, when the prompt includes the term ``still,'' the generated samples cluster closely around the ground-truth ``still'' samples after PCA projection, indicating good motion consistency.

In this part, ground-truth ``still'' samples serve as an anchor. A model performs better when its cluster is closer to this anchor and more compact, reflecting reduced redundant motion. StepVideo~\cite{ma2025step} and Hunyuan~\cite{kong2024hunyuanvideo} are closest to the anchor and form tight clusters, indicating strong frame-to-frame coherence. In contrast, VideoCrafter~\cite{chen2024videocrafter2overcomingdatalimitations} and VGen~\cite{zhang2023i2vgenxl} show more dispersed clusters, suggesting larger stochastic motion that deviates from the ``still'' ground truth. These findings align with~\autoref{fig:model_generation_vis}, where videos from VideoCrafter show inconsistent attention across frames during generation.



\subsubsection{Frequency Fluctuation} \label{freq_analsis} 
In~\autoref{fig:frequency}, we present the frequency spectra of real videos and those generated by several video generation models. Compared to real videos, all diffusion-based generators produce a more scattered high-frequency spectrum. For example, Hunyuan~\cite{kong2024hunyuanvideo} and LaVie~\cite{wang2023lavie} show ring-shaped energy bands, while StepVideo~\cite{ma2025step} produces sparse spikes along the axes. In contrast, VideoCrafter~\cite{chen2024videocrafter2overcomingdatalimitations} concentrates most energy in the low-frequency range, suggesting a loss of fine spatial details. These patterns highlight the difficulty for current VGM models to generate high-frequency details in a temporally consistent way.

\begin{figure*}[t]
    \centering
    \begin{subfigure}{0.3\textwidth}
    \caption*{X-CLIP}
        \includegraphics[width=\textwidth]{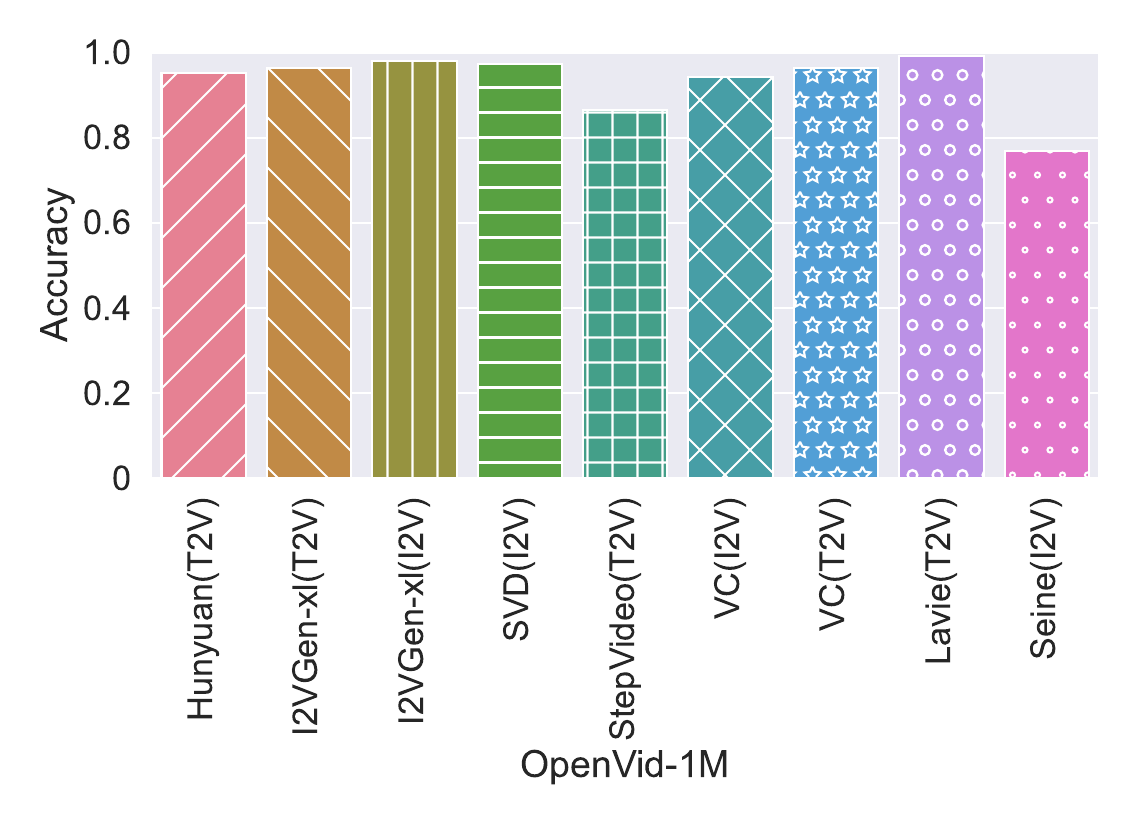}
    \end{subfigure}
    \begin{subfigure}{0.3\textwidth}
    \caption*{Masked Autoencoders}
        \includegraphics[width=\textwidth]{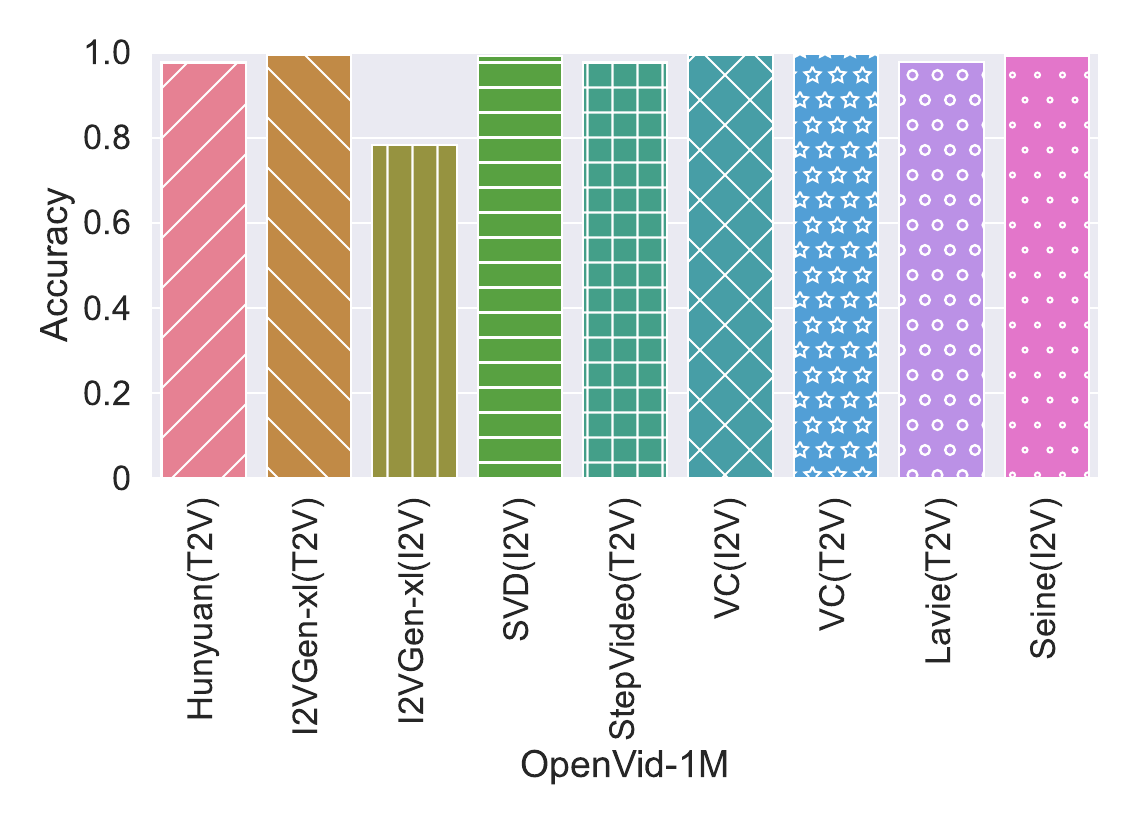}
    \end{subfigure}
    \begin{subfigure}{0.3\textwidth}
    \caption*{Inflated 3D CNN}
        \includegraphics[width=\textwidth]{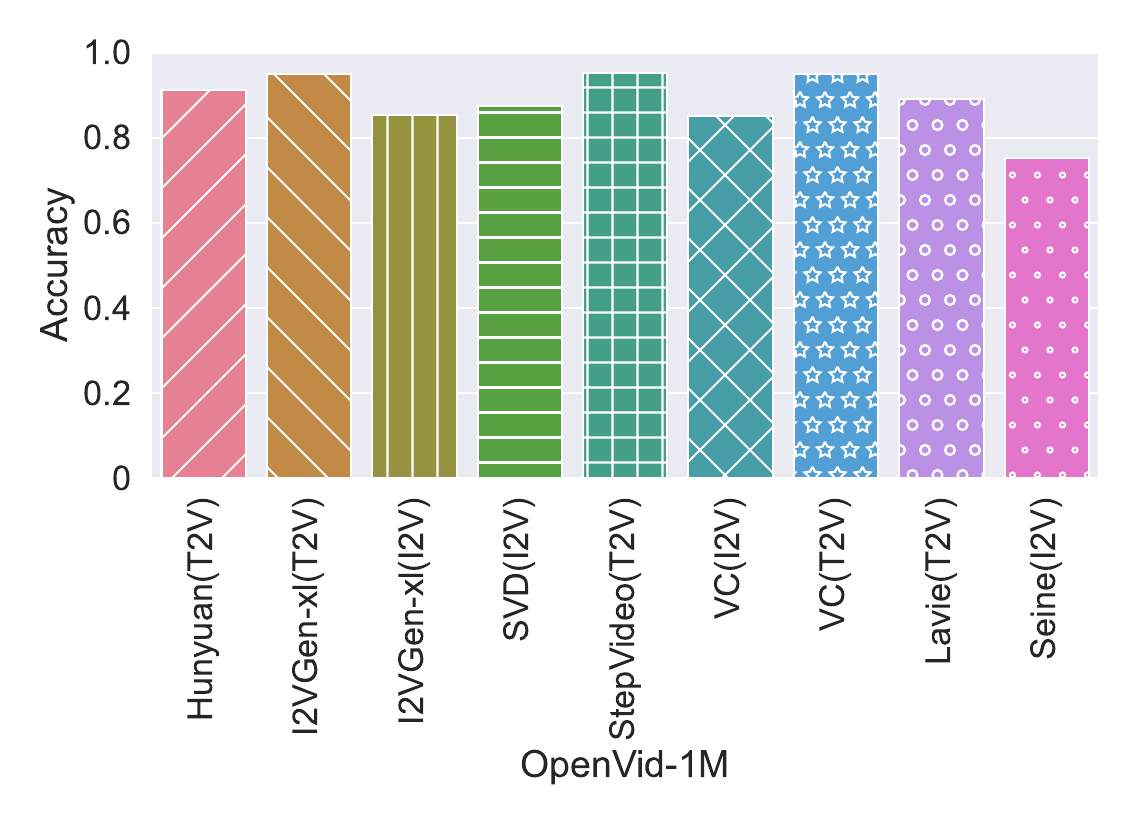}
    \end{subfigure}
    \\
    \begin{subfigure}{0.3\textwidth}
        \includegraphics[width=\textwidth]{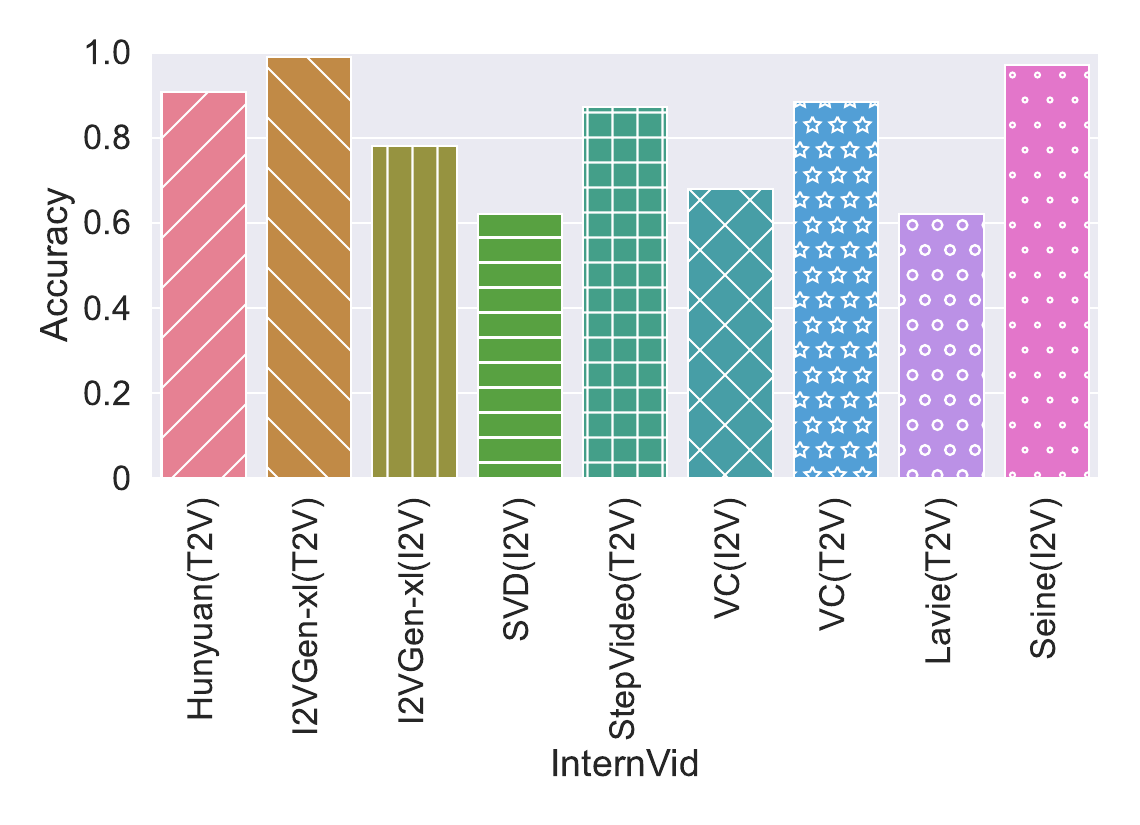}
    \end{subfigure}
    \begin{subfigure}{0.3\textwidth}
        \includegraphics[width=\textwidth]{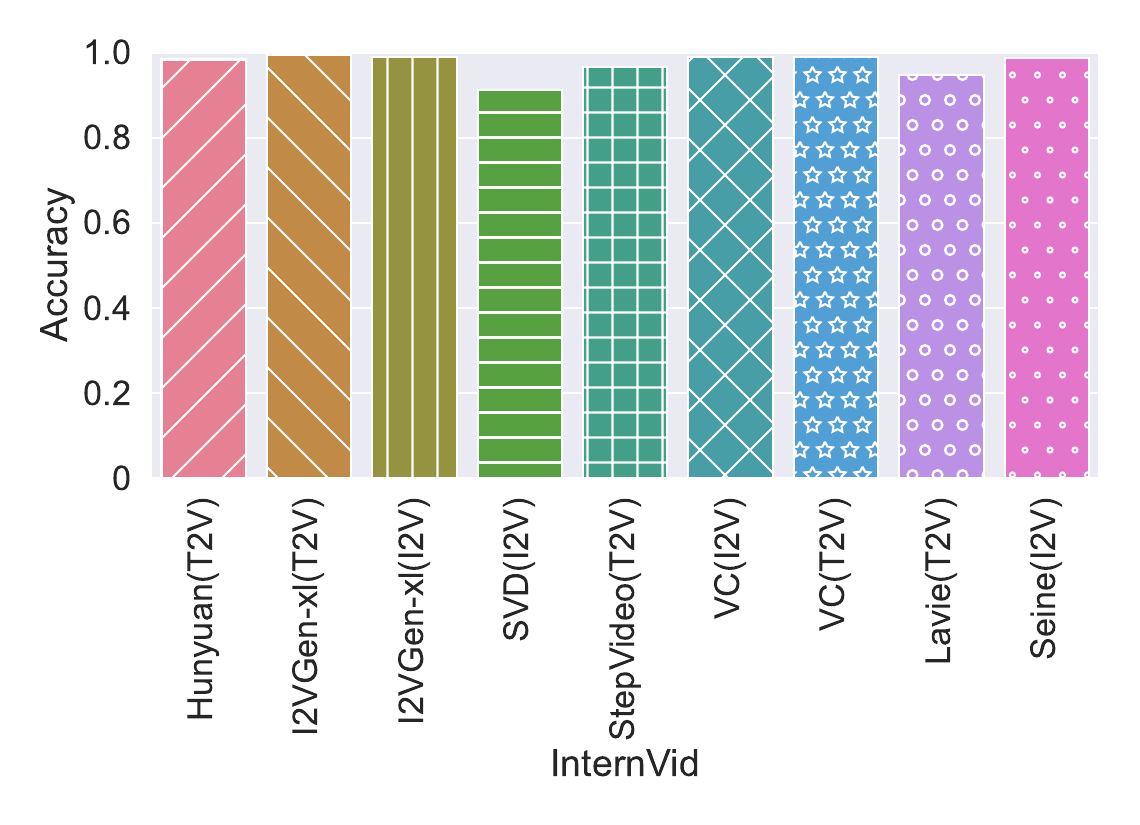}
    \end{subfigure}
    \begin{subfigure}{0.3\textwidth}
        \includegraphics[width=\textwidth]{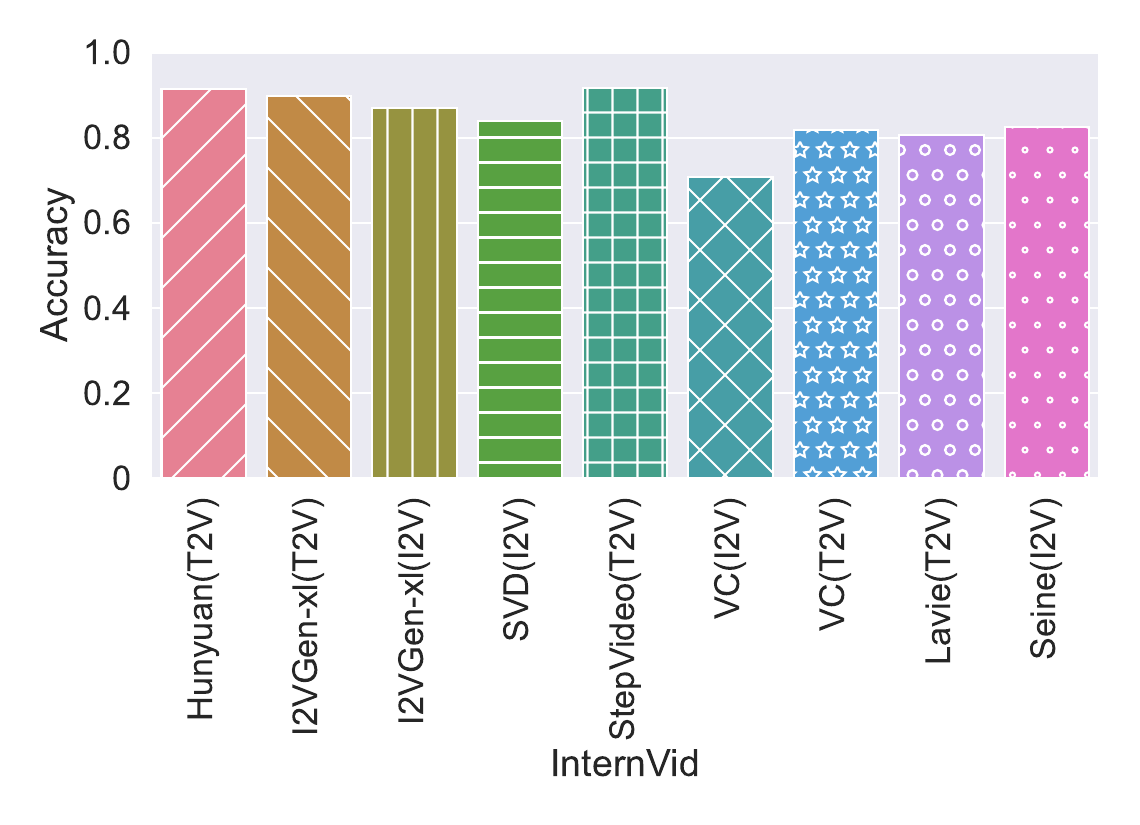}
    \end{subfigure}
    \caption{From left to right, the models utilize \xclip~\cite{ma2022xclip}, \mae~\cite{tong2022videomae}, and \id~\cite{carreira2018quo} as backbones for constructing detection models. The first row presents detection results for synthesized videos using data from OpenVid~\cite{nan2025openvid1mlargescalehighqualitydataset}, while the second row features videos generated with the InternVid~\cite{wang2024internvid}.}
    \label{fg:detection_video}
\end{figure*}

\subsection{Evaluation} 
\label{sec:detection_eval}
\subsubsection{Experiment Setup} \label{sec:setup}
\ 
\vspace{0.5em}
\newline
\noindent\textbf{Datasets.} We consider two datasets, both of which are traditional video-caption datasets that have been extensively used in training and evaluation on these models~\cite{wang2023videofactory,singer2022makeavideo,wang2023lavie,chen2024videocrafter2overcomingdatalimitations}.
\begin{itemize}[leftmargin=*]
\item \textbf{OpenVid-1M~\cite{nan2025openvid1mlargescalehighqualitydataset}.} OpenVid-1M is a significant, high quality dataset for test-to-video task. As a precisely curated collection, it contains over $1$ million video clips, totaling $2,051$ hours, including $0.4$ million $1080P$ resolution videos. Each clip is paired with expressive and detailed captions generated by large multimodal models, making it an effective resource for training advanced video generation models. OpenVid-1M offers carefully filtered content, selected for aesthetics, clarity, motion, and temporal consistency.
\item 
\textbf{InternVid~\cite{wang2024internvid}.}
InternVid is a large-scale, video-centric multimodal dataset. It is designed to facilitate learning robust and transferable video-text representations, crucial for multimodal understanding and generation tasks. Including over $7$ million videos with a cumulative duration of nearly $760,000$ hours, the dataset offers $234$ million video clips. Each clip is coupled with text descriptions, totaling approximately $4.1$ billion words. 
\end{itemize}


\paragraph{Evaluation Setting.} As shown in~\autoref{tab:video_generative_model}, we have collected seven the open-source video generation models, including \texttt{VGen}~\cite{zhang2023i2vgenxl}, \texttt{VideoCrafter}~\cite{chen2024videocrafter2overcomingdatalimitations}, \texttt{Hunyuan} \cite{kong2024hunyuanvideo}, \texttt{StepVideo} \cite{ma2025step}, \texttt{Stable Video Diffusion (SVD)} \cite{blattmann2023stable}, \texttt{Lavie}~\cite{wang2023lavie}, \texttt{Seine}~\cite{chen2023seine}. These models can generate videos based on conditional text~\cite{kong2024hunyuanvideo,ma2025step,chen2024videocrafter2overcomingdatalimitations,zhang2023i2vgenxl} or images~\cite{zhang2023i2vgenxl,blattmann2023stable,chen2024videocrafter2overcomingdatalimitations,chen2023seine}. Specifically, \texttt{VGen} and \texttt{VideoCrafter} can take both images and prompts to synthesize videos. Therefore, we have nine video generation tasks, and for each task, we generate $1000$ fake videos using each dataset. We want to clarify here that there is no data overlap between each generation task. Due to the need to change code that fit our task, we conducted our detection and source tracing tasks on these nine generation tasks. The two closed-source models were used to test the robustness of \textit{misuse prevention}.

OpenVid~\cite{nan2025openvid1mlargescalehighqualitydataset} and InternVid~\cite{wang2024internvid} are both video-caption datasets and do not include image data. Hence, for image-to-video generation models, we will clip the first frame of the video to use as the image input that queries the model. The data used to train detection and tracing models is a 50-50 split between generated and real videos. More experiment details for our detection and source tracing models are represented in~\autoref{tab:experiment_detail}.

\paragraph{Video Recognition Models.}In this work, we used Inflated 3D ConvNet~\cite{carreira2018quo}, Video Masked Autoencoders~\cite{tong2022videomae}, and X-CLIP~\cite{ma2022xclip} to build our detection and source tracing model.

\id\footnote{\url{https://github.com/v-iashin/video_features}} is a convolution-based neural network. Specifically, \id incorporates a convolution kernel to learn from the temporal dimension~\cite{ji20123d}. 
\xclip\footnote{\url{https://github.com/microsoft/VideoX}} directly utilizes the pre-trained \texttt{CLIP}~\cite{radford2021learning} model for video recognition tasks, leveraging its cross-frame attention mechanism to share information across frames. 
Besides, \mae\footnote{\url{https://github.com/microsoft/VideoX}} extends image autoencoders~\cite{he2021masked} to the video domain. It employs temporal downsampling, cube embedding, and tube masking techniques to devise a novel masked approach. 
When applied to self-supervised learning by masking multiple frames' patches, this approach prevents the model from merely learning simple temporal correlations. Since both \i3d and \mae can process $16$ frames at a time, we apply \xclip twice on the same sample, each time using $8$ frames, to match the $16$-frame detection length.
All of these models have been adapted from their original code repositories. Similarly, Grad-CAM\footnote{\url{https://github.com/facebookresearch/SlowFast}} has also been appropriately modified for use in our tasks to assist in analysis.

\begin{table}[t]
    \caption{Experiment details.}
    \label{tab:experiment_detail}
    \centering
    \resizebox{0.47\textwidth}{!}{
    \centering
    \begin{tabular}{cccc}
    \toprule
         Parameters & \id & \xclip & \mae \\
         \midrule
         Input frame  & $16$ & $8$ & $16$ \\
         Training epoch & $20$ & $20$ & $20$ \\
         Learning rate & $10^{-4}$ & $10^{-4}$ & $10^{-4}$ \\
         Optimizer & Adam & Adam & Adam \\
         Resolution & $224 \times 224$ & $224 \times 224$ & $224 \times 224$ \\
         Warmup steps & $1000$ & $1000$ & $1000$ \\
         Detection run time (seconds)  & $\approx 689$ & $\approx 10700$  & $\approx 3700$  \\
         Tracing run time (seconds)  & $\approx 2907$ & $\approx 48353$  & $\approx 16043$  \\
    \bottomrule  
    \end{tabular}}
    
\end{table}

    

\subsubsection{Detection}

\ 
\vspace{0.5em}
\newline
\paragraph{Targeted Detection.} We conducted targeted detection across $9$ video generation tasks using $3$ detection models and $2$ datasets, resulting in $54$ detection experiments in total. The overall results are presented in~\autoref{fg:detection_video}, with detailed FPR and FNR metrics listed in~\autoref{tab:appendix_fake_video_openvid} and~\autoref{tab:appendix_fake_video_internvid}. Among the three detection models, \xclip~\cite{ma2022xclip} is limited to processing $8$ video frames due to its pretraining setup and parameter constraints, while \id~\cite{carreira2018quo} and \mae~\cite{tong2022videomae} support $16$-frame inputs. Focusing on the video sample using the OpenVid dataset~\cite{nan2025openvid1mlargescalehighqualitydataset}, we observe that all three detection models achieve over $90\%$ detection success rates. Only in certain tasks, such as \xclip on OpenVid-Seine-I2V~\cite{chen2023seine}, \mae on OpenVid-VGen-I2V, and \id on Openvid-VGen-I2V~\cite{zhang2023i2vgenxl}, do the accuracy rates drop below $80\%$.
In contrast, detection performance degrades notably on InternVid~\cite{wang2024internvid}. \xclip falls below $70\%$ on several tasks, including those involving SVD-I2V, VideoCrafter-I2V, and LaVie-T2V. \id also shows reduced accuracy, though less severe. Meanwhile, \mae maintains high accuracy (often over $90\%$) across most tasks, likely due to its cross-frame attention mechanism, which captures richer temporal features than the 3D CNN-based \id. The relatively small parameter size of \id may further constrain its performance in high-resolution I2V scenarios.

Overall, \mae~\cite{tong2022videomae} demonstrates the most accurate and robust detection results. These findings suggest that detection effectiveness correlates with a model’s capacity to capture temporal dependencies. While \xclip is bottlenecked by frame limitations and \id shows inconsistency, \mae consistently delivers reliable performance across both datasets.



\begin{table}[t]
    \centering
    \caption{\mae-based \textit{detection} on four settings using the InternVid dataset~\cite{wang2024internvid}.}
    \label{tab:untargeted_detection}
    \Huge
    \resizebox{0.47\textwidth}{!}{
    \begin{tabular}{lcccc}
    \toprule[2.5pt]
         Scenario & Targeted detection & M-blind & D-blind & Open detection \\
         \midrule[1.5pt]
         HunyuanVideo~\cite{kong2024hunyuanvideo} (Text2Video)& $0.98$ & $0.71$ & $0.81$ & $0.81$\\
         VGen~\cite{zhang2023i2vgenxl} (Text2Video)&$0.99$ & $0.94$ & $0.99$  & $0.99$\\
         VGen~\cite{zhang2023i2vgenxl} (Image2Video)& $0.99$& $-$ & $0.82$ & $0.79$\\
         SVD~\cite{blattmann2023stable} (Image2Video)& $0.91$ & $0.59$ & $0.71$  & $0.59$\\
         StepVideo~\cite{ma2025step} (Text2Video)& $0.97$ & $0.72$ & $0.86$ & $0.54$\\
         VC~\cite{chen2024videocrafter2overcomingdatalimitations} (Image2Video) & $0.99$ & $0.65$ & $0.80$ & $0.98$\\
         VC~\cite{chen2024videocrafter2overcomingdatalimitations} (Text2Video) & $0.99$ & $0.62$ & $0.92$ & $0.92$\\
         Lavie~\cite{wang2023lavie} (Text2Video)& $0.95$ & $0.58$ & $0.62$  & $0.80$ \\
         Seine~\cite{chen2023seine} (Image2Video)& $0.99$ & $0.89$ & $0.97$ & $0.99$\\
         \bottomrule[2pt]
    \end{tabular}}
\end{table}

\paragraph{Untargeted Detection.} We then focus on utilizing \mae-based detection model evaluate on InternVid dataset~\cite{wang2024internvid} and show results in~\autoref{tab:untargeted_detection}. We observe that the detection model trained on fake videos generated by \texttt{VGen} (I2V)~\cite{zhang2023i2vgenxl} exhibits the highest accuracy in \textit{targeted detection}. Thus, this detector is utilized in \textit{M-blind} setting to individually assess videos from other generative tasks. It is distinctly noted that the detection model's effectiveness significantly declines in tasks such as \texttt{SVD} (I2V)~\cite{blattmann2023stable}, \texttt{VC} (I2V)~\cite{chen2024videocrafter2overcomingdatalimitations}, \texttt{VC} (T2V)~\cite{chen2024videocrafter2overcomingdatalimitations}, and \texttt{Lavie} (T2V)~\cite{wang2023lavie}. With a notable drop of up to $37\%$ in \texttt{VC} (T2V) and \texttt{Lavie} (T2V). This suggests that the video characteristics inherent in the fake videos produced by \texttt{VGen} (I2V) differ from those generated by these tasks. Furthermore, the \mae-based detector effectively identifies fake videos generated by other tasks. This ability to discern model `patterns' in generated videos underscores \mae's potential in fulfilling real-world detection tasks.

\textit{D-blind} involves using detection models trained on fake videos generated from the same models but querying from another dataset. While the detection accuracy of some tasks, such as \texttt{SVD} (I2V)~\cite{blattmann2023stable} and \texttt{Lavie} (T2V)~\cite{wang2023lavie}, shows a decline, the overall accuracy remains significantly higher compared to \textit{M-blind}. From the results presented in~\autoref{tab:untargeted_detection}, we observe that unknown generative models pose a more significant challenge compared with \textit{D-blind}.


For \textit{open detection}, we adopt the leave-one-out approach (according to~\autoref{sec:detection_meth_detect}) across all models. Specifically, we leave out one task's InternVid-generated videos as the test set, while using OpenVid-generated videos from remaining tasks as the training set. This setup reveals which models rely most on data-source uniformity and model consistency.

As shown in~\autoref{tab:untargeted_detection}, applying \textit{open detection} generally reduces accuracy, though in some tasks it even improves. This suggests that fake videos used for training share features with those from the left-out model. Thus, even without the left model’s specific patterns, the detector remains effective. However, \texttt{SVD} achieves only $59\%$ accuracy, likely due to its unique generation patterns not present in other models’ outputs, leading to the lowest performance in this setting.

\begin{figure}[!t]
    \centering
    \includegraphics[width=0.8\linewidth]{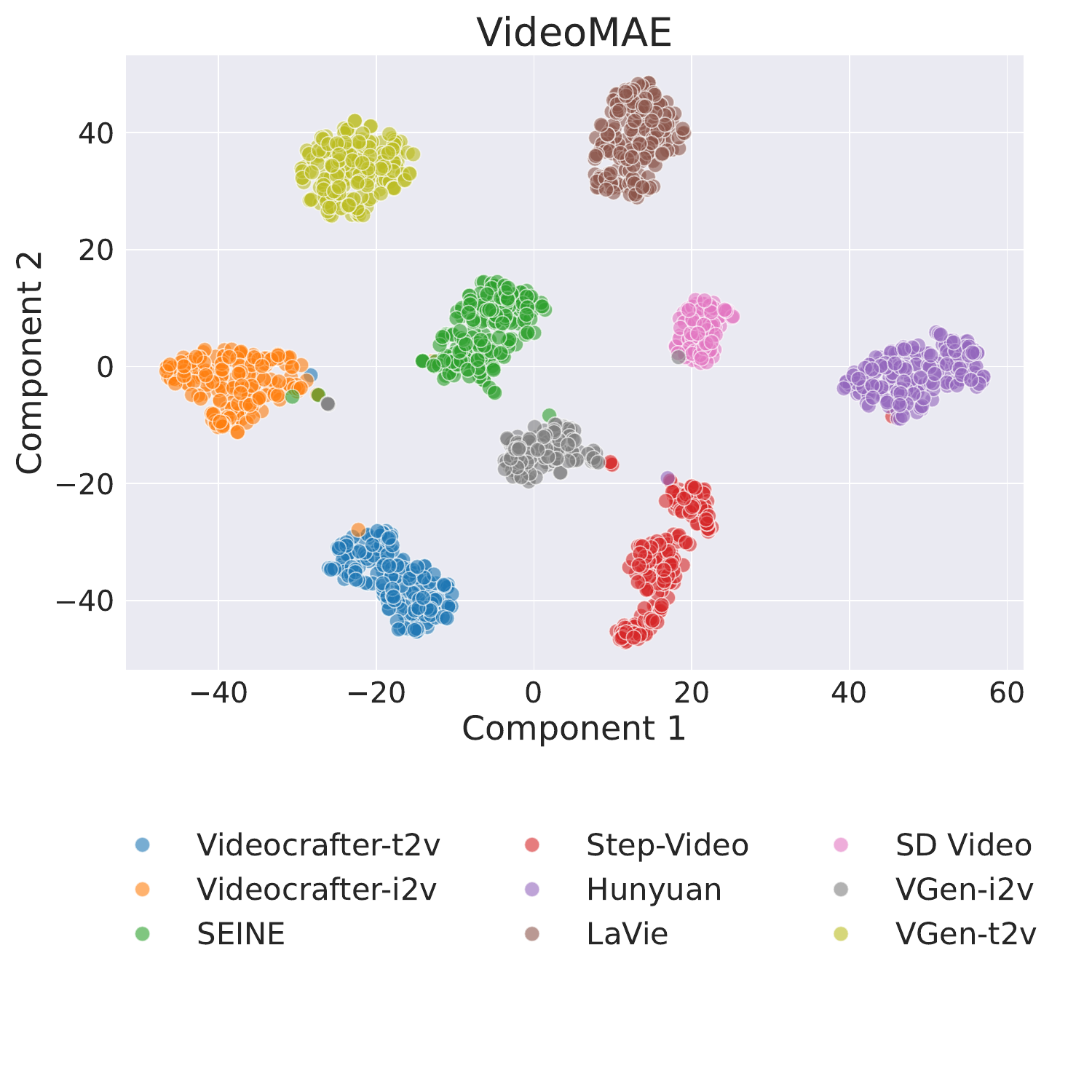}
    \vspace{-8mm}
    \caption{Presenting \mae-extracted features from videos generated by different VGMs. The t-SNE visualization shows distinct clusters by source.}
    \label{fig:tsne-st}
\end{figure}
\begin{tcolorbox}[breakable, colback=takeaways, boxrule=0pt]
\textit{Takeaways}: \textit{Targeted detection} proves the model's proficiency in accurately recognizing fake videos. \textit{D-blind} results show that fake videos generated by the same model but with different datasets share detectable `patterns'. \textit{M-blind} findings reveal that videos from different models but similar data sources possess distinguishable features. Lastly, \textit{open detection} demonstrates our model's effectiveness across all video generation models in data-independent and model-agnostic scenarios. It can accurately identify fake videos with sufficient training data.
\end{tcolorbox}

\begin{figure}[t]
    \centering
    \includegraphics[width=0.5\textwidth]{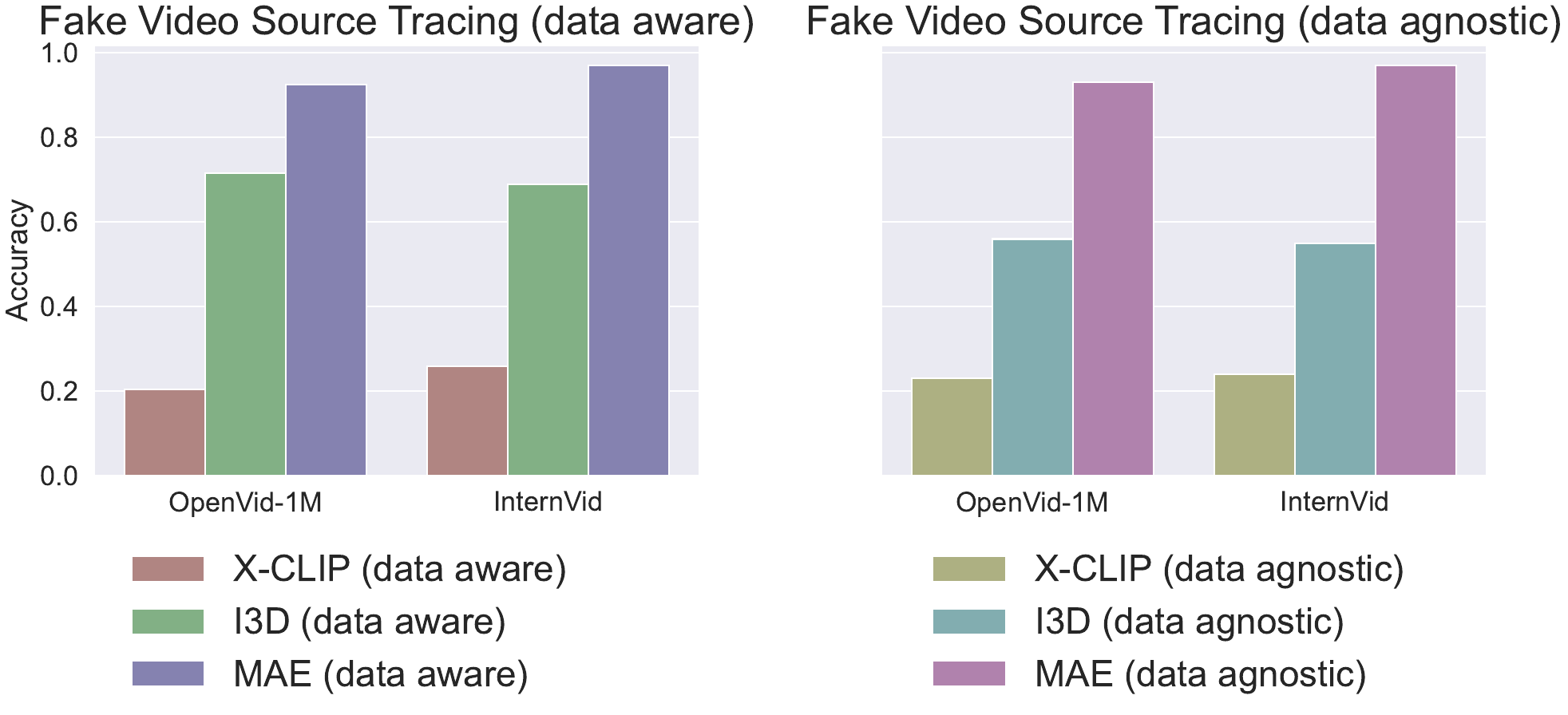}
    \caption{The results of source tracing under \textit{data-aware} and \textit{data-agnostic} settings on Openvid and InternVid datasets.}
    \label{fig:source_tracing}
\end{figure}

\subsubsection{Source Tracing}
\ 
\vspace{0.5em}
\newline
\noindent\paragraph{Data-Aware Setting.}In the \textit{data-aware} source tracing task, the goal is to assess whether generative models leave detectable patterns in the videos they produce. Our results suggest that model performance is highly sensitive to the temporal length of input videos. Models like \xclip~\cite{ma2022xclip}, which process limited frames, struggle to capture sufficient temporal patterns, leading to near-random accuracy. In contrast, \id~\cite{carreira2018quo} benefits from longer temporal windows, achieving moderate improvements. Notably, \mae~\cite{tong2022videomae} consistently delivers superior and stable results, likely due to its stronger capacity to model spatial-temporal dynamics.

As shown in~\autoref{fig:tsne-st}, features extracted by \mae form well-separated clusters across different generation models. This supports the idea that each model introduces unique spatial-temporal patterns, which \mae successfully captures as discriminative signatures.

\paragraph{Data-Agnostic Setting.}
Even in scenarios where the data source of the generated videos is uncertain, the \mae-based source tracing model still achieves an accuracy of $90\%$. Although this represents a slight decline from the source tracing accuracy in \textit{data-aware} setting, it is less than a $10\%$ decrease. This performance still significantly surpasses the \xclip- and \id-based source tracing models, which have detection accuracy of only $20\%$ and $50\%$, respectively.

We attribute this to \mae's ability to discern the distinct `patterns' carried by videos generated from different video generation tasks, as proposed in~\autoref{sec:source_tracing_methodology}. Therefore, the \mae-based source tracing model is able to perform tracing tasks of fake videos in an open-world context.

\begin{tcolorbox}[breakable, colback=takeaways, boxrule=0pt]
\textit{Takeaways}: Our experimental results demonstrate that the source tracing model using \mae as its backbone exhibits superior performance, achieving a $90\%$ accuracy rate in source tracing under agnostic data conditions. Our designed model can trace the source accurately using the generative model's features on its generated videos, without considering the data source.
\end{tcolorbox}

\subsection{Adaptive Attack}

The previous sections demonstrated the fantastic accuracy of \textit{detection} and \textit{source tracing}. However, \mg might use several generative models as a pipeline to generate a fake video. Therefore, in this section, we want to test our detection and source tracing models' performance under multimodel-generated fake video tasks. Because our work mainly focuses on the text-to-video and image-to-video generation models, we assume the \mg will first use a prompt to query one model and get the fake video. Then, use the first frame of the generated video and another prompt to get synthesized from the second model. Specifically, the selected models for this section need to be able to do the text-to-video and image-to-video tasks. Thus, we chose VGen~\cite{zhang2023i2vgenxl} and VideoCrafter~\cite{chen2024videocrafter2overcomingdatalimitations} in our work. 

To better examine the `patterns' left in the generated videos, we designed the prompt to follow the format: ``Two [object] are [description], the [left/right] is [description].'' For example, if we give ``Two trees in a serene meadow, the left tree is an ancient oak, majestic and tall'' to the first model. We will feed the second model with the first frame of generated video from the first model and counter prompt ``Two trees in a serene meadow, the right tree is a blossoming cherry, delicate and colorful.'' In our experiments, we created $30$ pairs of prompts in this format and presented our experiment results in~\autoref{tab:adaptive_attack}. Because the prompt is designed by ourselves, we employ \textit{D-blind detection} and \textit{Data-Agnostic source tracing} scenarios in this experiment. 

\begin{table}[t]
    \belowrulesep=0pt
    \aboverulesep=0pt
    \centering
    \caption{Detection and source tracing accuracy from videos generated from the multimodel pipeline. The test accuracy is all obtained from the final output of the generative pipeline. `First Model' represents the model that does the text-to-video task, and `Second Model' does the image-to-video task. For \textit{source tracing}, the accuracy should be the two numbers (first and second model) added together because we calculate the model probability separately.}
    \label{tab:adaptive_attack}
    \resizebox{0.47\textwidth}{!}{
    \begin{tabular}{c|ccc}
    \toprule
        \multicolumn{2}{c}{Tasks} & First Model & Second Model \\ 
        \midrule
        \multirow{2}{*}{Detect} & VGen-VC & $0.97$ & $0.99$ \\
        & VC-VGen & $0.67$ & $0.79$ \\
        \midrule
        \multirow{2}{*}{Source Tracing } & VGen-VC & $0.63$ & $0.30$ \\
        & VC-VGen & $0.38$ & $0.46$ \\
    \bottomrule
    \end{tabular}}
\end{table}


In~\autoref{tab:adaptive_attack}, we evaluate two multi-model pipelines: `VGen-VC' (VGen~\cite{zhang2023i2vgenxl} followed by VideoCrafter~\cite{chen2024videocrafter2overcomingdatalimitations}) and `VC-VGen' (the reverse order). Detection performance is case-sensitive: `VGen-VC' is easily detected, while `VC-VGen' yields only ~$70\%$ accuracy--lower than in the \textit{D-blind} setting~\autoref{tab:untargeted_detection}. Similarly, \textit{source tracing} accuracy drops significantly, identifying only ~$30\%$ of participating models in both pipelines. Since model attribution is computed per model, the total tracing score sums both predictions (e.g., $0.63+0.30$ for VGen-VC). These results indicate that both detection and tracing models degrade when handling videos generated by multi-model pipelines.

\begin{figure}
    \includegraphics[width = 0.5\textwidth]{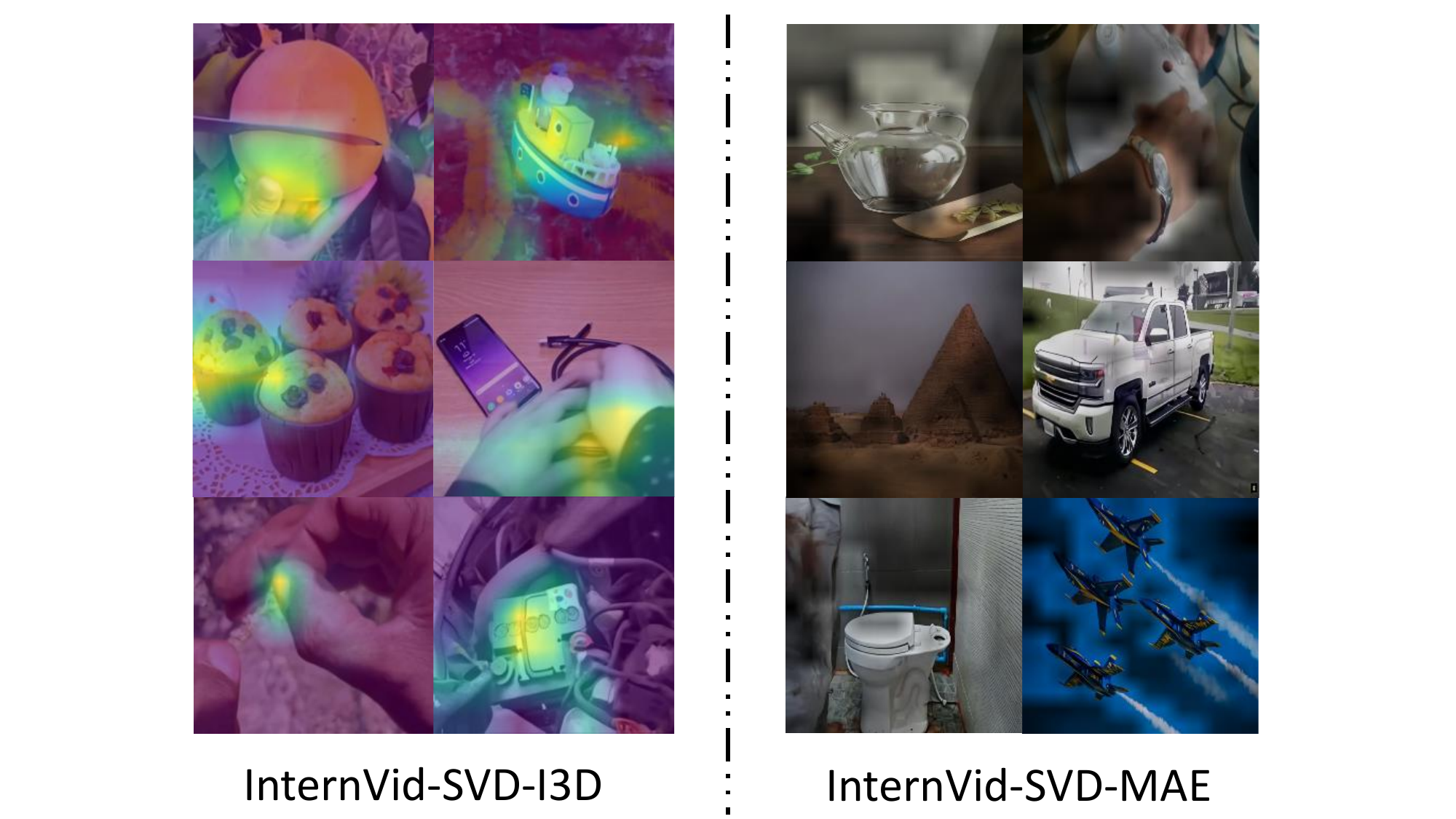}
    \caption{Grad-CAM~\cite{Selvaraju_2019} (left) and attention visualization~\cite{chefer2021transformer} (right) on the first frame of InternVid-generated videos~\cite{wang2024internvid}. The \id-based model focuses on localized pixels with static attention, while the \mae-based model adaptively attends to different anomalous objects across frames and captures temporal inconsistencies via motion and shape cues.}
    \label{fig:detection_ablation}
    \vspace{-1em}
\end{figure}

\subsection{Detailed Analysis} \label{sec:detailed_analysis}

\subsubsection{Detection}\label{sec:detection_ablation}

As shown in~\autoref{fg:detection_video}, both \id~\cite{carreira2018quo} and \mae~\cite{tong2022videomae}-based detection models achieve over $90\%$ accuracy. To understand this strong performance, we investigate the cues these models rely on, using Grad-CAM~\cite{Selvaraju_2019} for \id and attention map visualization~\cite{chefer2021transformer} for \mae.

\autoref{fig:detection_ablation} presents examples from InternVid-SVD, showing that both models focus on objects--e.g., boats, cakes, or watches--that often appear distorted in generated videos. These object-level anomalies reflect limitations of some video generation models in spatial reconstruction.

At the temporal level, we observe notable discontinuities in attention dynamics, which align with our findings of attention shifts in~\autoref{sec:preliminary_analysis}. The \id-based model maintains static focus across frames, often missing evolving inconsistencies. In contrast, the \mae-based model adapts its attention to moving objects, effectively capturing temporal distortions (e.g., planes in motion or shape-shifting plates).

These findings suggest that while both models detect spatial anomalies, only the \mae-based model reliably captures temporal inconsistencies. This explains its superior detection accuracy and highlights potential limitations of the \id-based model as generative models improve object fidelity.


\begin{tcolorbox}[breakable, colback=takeaways, boxrule=0pt]
\textit{Takeaways}: The \id-based detection model primarily relies on spatial distortions of objects to assess video authenticity. Its attention remains relatively static across frames, limiting its ability to capture temporal anomalies. In contrast, the \mae-based model is more adaptable, attending to both spatial irregularities and motion inconsistencies across time.

These findings align with our earlier analysis in~\autoref{sec:preliminary_analysis}, which identified attention shift, motion variation, and frequency spectra as key indicators of generated videos. The \mae-based model’s ability to track dynamic attention and detect subtle temporal changes allows it to better leverage these multimodal cues, leading to superior detection performance. 
\end{tcolorbox}

\subsubsection{Source Tracing} \label{sec:detect_analysis_st}

Following~\autoref{sec:detection_ablation}, we further examine which attributes the source tracing model relies on. We queried four models--\texttt{VGen}~\cite{zhang2023i2vgenxl}, \texttt{VideoCrafter}~\cite{chen2024videocrafter2overcomingdatalimitations}, \texttt{Stable Video Diffusion}~\cite{blattmann2023stable}, and \texttt{Seine}~\cite{chen2023seine}--using unseen prompts. Given the superior performance of the \mae-based model, we adopted it for all source tracing tasks.

After verifying correct attribution, we visualized attention maps to understand which video regions guided the model’s predictions. As shown in~\autoref{fig:source_tracing_ablation}, the model attends to different elements depending on the generator. \texttt{Stable Video Diffusion} generates videos with coherent motion and emission patterns, indicating strong temporal consistency. In contrast, outputs from \texttt{VGen} and \texttt{Seine} often exhibit noticeable shape distortions or emission inconsistencies. These anomalies are effectively captured by our MAE-based model, which operates in a data-agnostic manner. Rather than relying on generator-specific patterns, it detects generative artifacts directly from video content, demonstrating strong generalization across different models.

\begin{tcolorbox}[breakable, colback=takeaways, boxrule=0pt]
\textit{Takeaways}: Videos generated from the same image by different models will contain unique characteristics specific to each model. For example, characteristics such as deviations in the trajectories of moving objects and distortions in their shapes. These characteristics can assist the source tracing model in effectively tracing the origin of the videos. Moreover, these features are agnostic to data sources.
\end{tcolorbox}

\section{Misuse Prevention} \label{sec:misuse_prevention}




Compared to text-to-video generation tasks, videos produced through image-to-video generation tasks are more susceptible to abuse due to the existence of \mg. They are more likely to lead to copyright infringement and misuse issues. At the same time, with the development of video generation models, fake videos might be more time-coherent and have a higher resolution. Therefore, only employing \textit{detection} in such scenarios may not be enough. 
A dedicated defensive strategy is needed for this type of generation tasks. We plan to design a defense mechanism based on the concept of {\it adversarial examples}, tailored explicitly for the image-to-video generation task.

Adversarial examples are first introduced to target misclassification problems, incorporating a small perturbation to the original image but are invisible to humans~\cite{goodfellow2015explaining,papernot2015limitations,moosavidezfooli2016deepfool,kurakin2017adversarial,carlini2017towards,tramèr2017space,madry2019deep}. 
%
%

\begin{figure*}[h]
    \centering
    \includegraphics[width=\textwidth]{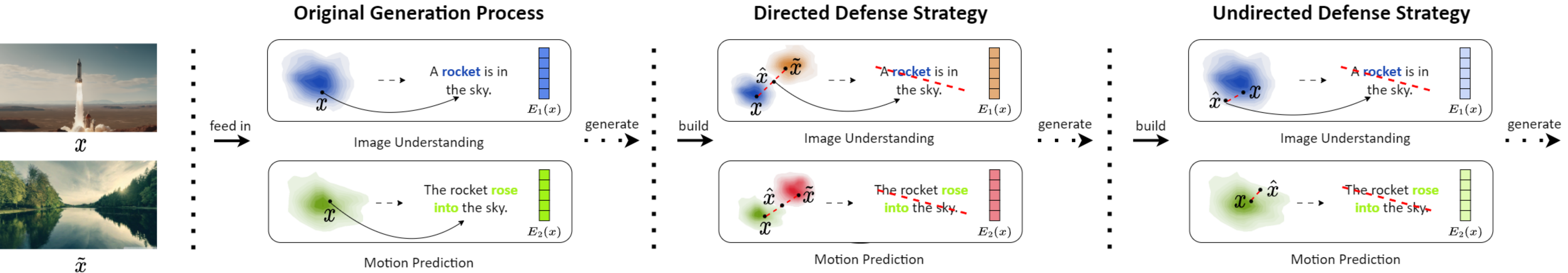}
    \caption{Prevention strategies are implemented by introducing perturbations to $x$, causing semantic shifts. \textit{Directed defense} employs a selectively chosen $\tilde{x}$ for guidance, while \textit{undirected defense} adds perturbations indiscriminately.}
    \label{fig:ad_defense}
\end{figure*}

Szegedy et al.~\cite{szegedy2014intriguing} first proposed the concept of adversarial samples. Their work narrates how adding perturbations to an image can lead to incorrect judgments by neural networks. Following this, methods for generating adversarial examples, such as FGSM~\cite{goodfellow2015explaining}, C\&W~\cite{carlini2017towards}, and PGD~\cite{madry2019deep}, emerged.
%
Early adversarial examples were primarily used in classification models, aiming to confuse classifiers to elicit incorrect categorization results. 
In our task, we aim to confuse video generative models, and hence, we employ a method similar to PGD~\cite{madry2019deep} to create adversarial examples for defense.


Our intuition is video generation models utilize encoders to analyze objects within the input images. For instance, as demonstrated in~\autoref{fig:ad_defense}, if the image contains a rocket, the model discerns the object in the image and generates corresponding continuous frames, ultimately stringing these frames together to create a video. This setting differs from prior works in the image domain, as our model is required to predict object motion in videos rather than merely recognize it. Thus, it is necessary to deceive the model's semantic and motion prediction encoders into misinterpreting the image, thereby producing incorrect and bizarre frames, ultimately safeguarding the image from misuse.


\subsection{Methodology}

We use $E_{1}(\cdot)$ to denote the model's understanding of the image in the spatial domain, and $E_{2}(\cdot)$ for its understanding in the temporal domain. We introduced a directed defense approach and an undirected one; they are different in that the directed approach needs the \oc to pick a target image $\tilde{x}$ and the undirected does not. We will provide a detailed discussion of these two methods in the following.

\paragraph{Directed Defense.} The method of \textit{directed defense} involves using a guiding image to direct the perturbations added to $x$, which we refer to as the target image $\tilde{x}$. Our aim is for the modified image $\hat{x}$ to be similar to the original image $x$ at the pixel level while resembling $\tilde{x}$ at the semantic level. Accordingly, we have crafted our optimization objective as follows:
\begin{align}
    \underset{\hat{x}}{\mathrm{arg\,min}}\;\;  &\| E_1(\hat{x}) - E_1(\tilde{x})\|_{\ell_1} + \lambda_1 \cdot \|E_2(\hat{x}) - E_2(\tilde{x})\|_{\ell_1} \nonumber \\  
    &+ \lambda_2 \cdot \left[\| \hat{x} - x \|_{\ell_2} +L_{\text{lpips}}(\hat{x},x)\right] 
    \label{eq:objective_directed}
\end{align}
Herein, we desire the generated adversarial example $\hat{x}$ to attain a similar semantic understanding when processed by the $E_1$ and $E_2$ encoders. This calculation will employ either the $L_1$ norm or cosine similarity matrix. Concurrently, we use the $L_2$ norm and $L_{\text{lpips}}$ loss between $\hat{x}$ and $x$ to ensure similarity at the pixel level.

\begin{algorithm}
    \caption{Directed Defense}
    \label{alg:directed_defense}
    \begin{algorithmic}[1]
    \Require Original image $x$, target image $\tilde{x}$, image encoder $E_1$, video encoder $E_2$, optimization rate $\mu$, upper bound $\eta$, number of iterations $T$
    \State Utilizing objection function as defined in~\autoref{eq:objective_directed}
    \State Set the initial adversarial example $\hat{x}_{0}$ $\leftarrow$ $x$ 
    \For{$i\gets 0$ \textbf{to} $T-1$} \Comment{Perform $T$ repetitive iterations.}
    \State $\hat{x}^{*}_{i} \leftarrow \hat{x}_{i} - \mu \cdot sgn (\nabla_{\hat{x}_{i}}L(\hat{x}_{i},\tilde{x},E_1,E_2))$ 
    \State $\beta \leftarrow \hat{x}^{*}_{i} - x$ and bound $\beta \leq \| \eta \|_{\ell_1}$
    \State $\hat{x}_{i+1} \leftarrow x + \beta $
    \EndFor
    \Ensure $\hat{x}_{T}$  \Comment{$\hat{x}_{T}$ simplify denoted as $\hat{x}$ in our paper}
    \end{algorithmic}
\end{algorithm}
Theoretically, we can take any off-the-shelf optimizer to find $\hat{x}$. In our setting, we apply a PGD-style method, as shown in~\autoref{alg:directed_defense}.  Specifically, we compute the loss for each iteration using the loss function from~\autoref{eq:objective_directed}, and after calculating the derivative, we subtract this gradient value from the current iteration's $\hat{x}^{i}$. This is because our \textit{directed defense} is formulated as an optimization problem aimed at approximating the target image's projections $E_1(\cdot)$ and $E_2(\cdot)$, thereby necessitating the application of gradient descent. We treat $\eta$ as hyperparameters in our experiments, and we will evaluate them in~\autoref{sec:prevention_eva}.

\paragraph{Undirected Defense.} The target image $\tilde{x}$ substantially influences the efficacy of \textit{directed defense}. Careful selection of the target image is imperative to achieve optimal defensive performance. However, this selection process often necessitates semantic and pixel filtering, which varies depending on the original image. To obviate the laborious task of selecting a proper target image for each unique original image, we propose our \textit{undirected defense} method. This allows for the implementation of defense strategies irrespective of the original image.
\begin{align}
    \underset{\hat{x}}{\mathrm{arg\,max}}\;\; & \| E_1(\hat{x}) - E_1(x)\|_{\ell_1} \nonumber + \lambda_1 \cdot \|E_2(\hat{x}) - E_2(x)\|_{\ell_1} \\
    &- \lambda_2 \cdot \left[\| \hat{x} - x \|_{\ell_2} +L_\text{lpips}(\hat{x},x)\right] 
    \label{eq:objective_undirected}
\end{align}
We posit that the adversarial example $\hat{x}$ requires iterative modifications to increase its distance from $x$ in the latent space projected by $E_1(\cdot)$ and $E_2(\cdot)$. We employ the $L_1$ norm to measure the distance between embeddings and iteratively optimize $\hat{x}$. Similar to \textit{directed defense} strategies, our optimization process also aims to maintain proximity to the original image $x$. To this end, we opt for the use of $L_{\text{lpips}}$ and $L_2$ norm to control pixel-level similarity. The objective function of \textit{undirected defense} is defined in~\autoref{eq:objective_undirected}.

The primary distinction between \textit{directed defense} and \textit{undirected defense} lies in three aspects. First is eliminating the need for a target image $\tilde{x}$. Second, the optimization objective is different. Third, a modification occurs in the computation sign at line $4$ in~\autoref{alg:directed_defense}. The \textit{undirected defense} transforms the method into a maximization optimization problem. Similar to the traditional methods of generating adversarial example~\cite{goodfellow2015explaining}, our objective is to add perturbations that amplify the loss in $E_{1}(\cdot)$ and $E_{2}(\cdot)$, effectively achieving a gradient ascent, hence the utilization of the addition sign. 

\begin{figure*}[h]
    \includegraphics[width=\textwidth]{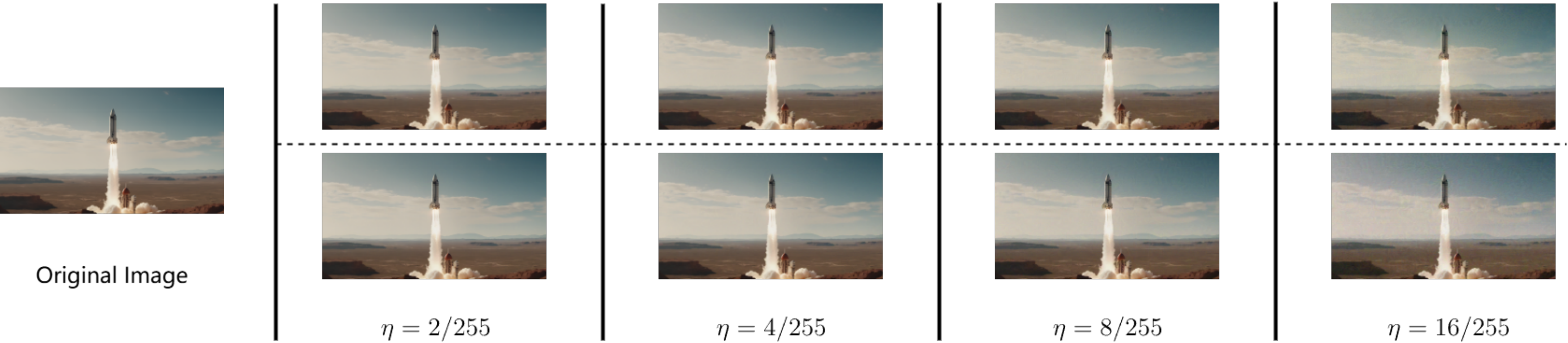}
    \caption{Adversarial examples with $\eta = \frac{2}{255}, \frac{4}{255}, \frac{8}{255}, \text{ and } \frac{16}{255}$, the first row applies a \textit{directed defense} method, and the second row an \textit{undirected defense} method.}
    \label{fig:defense}
\end{figure*}

\subsection{Evaluation} \label{sec:prevention_eva}
In this section, our experiments encompass not only open-source models capable of supporting image-to-video generation, such as \texttt{SVD}~\cite{blattmann2023stable}, but we will also conduct tests on several commercial models referenced in ~\autoref{sec:background_diffusion} to further substantiate the effectiveness of our approach.

We will adopt the original adversarial strategy against image generation as the baseline method, which primarily focuses on object understanding and overlooks motion prediction in the image. We use two embedders from \texttt{SVD}~\cite{blattmann2023stable} in our experiments. $E_{1}$, the first-layer embedder, is the `FrozenOpenCLIPImagePredictionEmbedder'\footnote{CLIP image encoder server as the `FrozenOpenCLIPImagePredictionEmbedder' in Stable Video Diffusion.}. It generates embeddings for conditional frames. $E_{2}$, the fourth-layer embedder, is the `VideoPredictionEmbedderWithEncoder'\footnote{Stable Diffusion 2.1 encoder is utilized as the `VideoPredictionEmbedderWithEncoder' in Stable Video Diffusion.}. It creates inputs for UNet, aiming for temporal-level prediction. The parameters $\lambda_1$ and $\lambda_2$ are both set to $1$ in~\autoref{eq:objective_directed} and~\autoref{eq:objective_undirected}. $\mu$ is set to $\frac{1}{255}$, as this configuration has demonstrated effective defense capabilities.

\subsubsection{Directed Defense}\label{sec:is_eva_dd}

This section showcases adversarial examples generated using~\autoref{alg:directed_defense}, as shown in~\autoref{fig:defense}. We test four $\eta$ values: $\frac{2}{255}, \frac{4}{255}, \frac{8}{255}, \text{and } \frac{16}{255}$. Since~\autoref{alg:directed_defense} requires a target image $\tilde{x}$ as guidance, the output $\hat{x}$ inevitably inherits visual imprints from $\tilde{x}$. To ensure perturbations remain imperceptible, we cap $\eta$ at $\frac{16}{255}$.

As shown in~\autoref{fig:appendix_shield}, setting $\eta = \frac{4}{255}$ affects videos generated by Stable Video Diffusion~\cite{blattmann2023stable}. The rocket stays suspended mid-air while the background clouds move, revealing a temporal inconsistency. Compared to the baseline method, which adds perturbations only at the semantic level, our approach at $\eta = \frac{4}{255}$ better disrupts object motion. The baseline still allows the model to recognize objects and generate plausible motion.

We further evaluate our $\eta = \frac{4}{255}$ example on other models in~\autoref{fig:appendix_compare}. On Gen-2, the adversarial input disrupts the rocket’s motion: although smoke is still emitted, the rocket remains stationary. For Pika Lab, which already produces subtle motion, major anomalies are not observed in the rocket, but the emitted smoke remains static post-launch.

\begin{table*}[!t]
    \belowrulesep=0pt
    \aboverulesep=0pt
    \caption{Utilize SSIM, PSNR, and LPIPS to examine the video quality after applying defense strategies.}
    \label{tab:prevention_eva}
    \centering
     \resizebox{1.0\textwidth}{!}{
    \begin{tabular}{c|c|ccc|ccc|ccc|ccc}
    \toprule 
         \multirow{2}{*}{Defense Method}&\multirow{2}{*}{Features}& \multicolumn{3}{c|}{ \rule{0pt}{3ex}\raisebox{0.5ex}{$\eta = \frac{2}{255}$}}&\multicolumn{3}{c|}{\rule{0pt}{3ex}\raisebox{0.5ex}{$\eta = \frac{4}{255}$}}&\multicolumn{3}{c|}{\rule{0pt}{3ex}\raisebox{0.5ex}{$\eta = \frac{8}{255}$}}&\multicolumn{3}{c|}{\rule{0pt}{3ex}\raisebox{0.5ex}{$\eta = \frac{16}{255}$}}\\ \cmidrule(lr){3-5} \cmidrule(lr){6-8} \cmidrule(lr){9-11} \cmidrule(lr){12-14}
         &  & SSIM & PSNR & LPIPS & SSIM & PSNR & LPIPS & SSIM & PSNR & LPIPS & SSIM & PSNR & LPIPS \\ \cline{1-14}
         \tt Baseline & spatial & \cellcolor{pink!25} ${0.339}$ & \cellcolor{pink!25} $10.52$ & \cellcolor{pink!25} $0.654$ & \cellcolor{pink!25} $0.341$ & \cellcolor{pink!25} $10.37$ & \cellcolor{pink!25} $0.669$ & \cellcolor{pink!25} $0.288$ & \cellcolor{pink!25} $10.58$ & \cellcolor{pink!25} $0.702$ & \cellcolor{pink!25} $0.301$ & \cellcolor{pink!25} $10.42$ & \cellcolor{pink!25} $0.696$ \\
         \tt Ours (directed) & \textbf{spatiotemporal} & \cellcolor{green!25} $\mathbf{0.314}$ & \cellcolor{green!25} $\mathbf{10.32}$ & 
         \cellcolor{green!25} $\mathbf{0.691}$ & \cellcolor{green!25} $\mathbf{0.336}$ & \cellcolor{green!25} $\mathbf{10.33}$ & \cellcolor{green!25} $\mathbf{0.674}$ & \cellcolor{green!25} $0.298$ & \cellcolor{green!25} $\mathbf{10.32}$ & \cellcolor{green!25} $0.695$ & \cellcolor{green!25} $0.301$ & \cellcolor{green!25} $\mathbf{10.37}$ & \cellcolor{green!25} $\mathbf{0.702}$ \\
         \tt Ours (undirected) & \textbf{spatiotemporal} & \cellcolor{green!25} $\mathbf{0.337}$ & \cellcolor{green!25} $10.54$ & \cellcolor{green!25} $\mathbf{0.67}$ & \cellcolor{green!25} $\mathbf{0.332}$ & \cellcolor{green!25} $\mathbf{10.27}$ & \cellcolor{green!25} $\mathbf{0.676}$ & \cellcolor{green!25} $0.331$ & \cellcolor{green!25} $\mathbf{10.45}$ & \cellcolor{green!25} $0.675$ & \cellcolor{green!25} $0.305$ & \cellcolor{green!25} $\mathbf{10.4}$ & \cellcolor{green!25} $\mathbf{0.714}$ \\
    \bottomrule
    \end{tabular}}
\end{table*}

\subsubsection{Undirected Defense}\label{sec:is_eva_ud}

The \textit{undirected defense} method does not require the guidance of a target image but instead directly optimizes the image to increase the loss. We observed in~\autoref{fig:defense} that at $\eta = \frac{16}{255}$, the adversarial example generated by the \textit{undirected defense} method appears more similar to the original image. This is compared to that generated by the \textit{directed defense} method, as seen in the first row. In the upper left portion of the sky in the two images, the \textit{directed defense} method introduces certain features from the target image. Meanwhile, the \textit{undirected defense} maintains more visual similarity and equivalent defense effectiveness. The images in~\autoref{fig:appendix_shield} show that the \textit{undirected defense} method is effective. Even at $\eta = \frac{4}{255}$, it can successfully prevent Stable Video Diffusion~\cite{blattmann2023stable} from generating reasonable videos. The rocket remains stationary in the air, which leads to the model's incorrect motion prediction. We also applied the baseline method using the \textit{undirected defense} procedure. Similar to findings in~\autoref{sec:is_eva_dd}, we found that merely disrupting the semantic understanding of the image without affecting object motion is insufficient for effective defense.

Similarly, we used adversarial examples generated by the \textit{undirected defense} as inputs for experimentation with Pika Lab and Gen-2. On Gen-2, we observed the rocket's motion, which did not significantly differ from the original video. However, the video generated from images processed by \textit{undirected defense} exhibited distorted and anomalous motion in the rocket's launch base, increasing the video's implausibility. For Pika Lab, generated videos resembled the originals, but the smoke after the rocket remained fixed in the sky, making the overall video appear slightly abnormal.

\subsection{Discussion}

To quantify the effectiveness of our defense methods, we evaluate the generated videos using three video quality metrics: SSIM, PSNR, and LPIPS. We exclude FVD due to its high sample requirements and computational cost. In this experiment, we select $30$ high-quality videos and compare our defense strategies with baseline methods (see~\autoref{tab:prevention_eva}).

As shown in~\autoref{tab:prevention_eva}, incorporating temporal features into the defense process reduces video quality. For instance, the \textit{directed defense} strategy yields an SSIM of $0.314$, lower than the baseline's $0.339$. Additionally, we apply the baseline, \textit{directed defense}, and \textit{undirected defense} strategies to Gen-2 and Pika Lab models, recording whether these defenses disrupt motion prediction during video generation (see~\autoref{tab:results_defense}).

Compared to \textit{undirected defense}, the \textit{directed defense} strategy is more effective. It works by guiding the image in latent space toward a target image, increasing the likelihood of misleading the video generation model. However, stronger perturbations (larger $\eta$) may introduce artifacts from the target image, reducing output quality. Hence, selecting a suitable target with similar scenes but distinct main objects is essential to maintaining visual quality and preventing motion misinterpretation.

\subsection{Limitation}

Although our experimental results show that these defense strategies can effectively disrupt the video generation process, their effectiveness is largely due to the fact that many current VGMs~\cite{chen2023seine, chen2024videocrafter2overcomingdatalimitations,blattmann2023stable,zhang2023i2vgenxl} still rely heavily on Stable Diffusion as the backbone. This leads to strong semantic similarities across models, making our SVD-based perturbations effective.

However, as models evolve, architectures are shifting toward alternatives like DiT, which are increasingly adopted in VGMs~\cite{kong2024hunyuanvideo,ma2025step}. This change reduces the effectiveness of our method against newer models.

Furthermore, since the core idea of our approach is to apply perturbations directly to the image, it remains vulnerable to image processing and purification techniques~\cite{hönig2025adversarialperturbationsreliablyprotect}--challenges also commonly faced by similar methods~\cite{liang2023adversarial,shan2023glaze} in the image domain. Therefore, identifying a stable and robust prevention strategy remains an open problem and a key focus of our future work.

\begin{tcolorbox}[breakable, colback=takeaways, boxrule=0pt]
\textit{Takeaways}: According to quantitative and visual evaluations, both directed and undirected defenses can effectively provide invisible protection to images. Moreover, \textit{directed defense} is effective against Stable Video Diffusion and causes disruption in videos generated using Gen-2 and Pika Lab models. We also acknowledge that as VGMs continue to evolve, if future models diverge significantly from \texttt{SVD} in both semantic and temporal patterns, the effectiveness of our defense may be greatly reduced. The defense is also vulnerable to image processing techniques.
\end{tcolorbox}

\section{Related Work} \label{sec:related_work}

\subsection{Fake Content Detection}

With the rise of generative models, concerns have grown around their misuse in creating fake visual and textual content. Prior works have focused on detecting fake images and videos~\cite{girish2021discovery,wang2020cnngenerated,yu2019attributing,sha2023defake,cozzolino2023synthetic,lu2023seeing,guera2018deepfake,wodajo2023deepfake}, as well as AI-generated text~\cite{krishna2023paraphrasing,mitchell2023detectgpt}. In image generation, efforts primarily aim to distinguish outputs from GANs and diffusion models~\cite{sha2023defake,yu2019attributing}. However, these methods are often domain-specific and fail to generalize to more complex video generation scenarios.

Existing detection approaches have largely focused on early-stage or low-resolution generation models~\cite{guera2018deepfake,dang2020detection,wang2020fakespotter}. For instance, datasets like FaceForensics++\cite{rossler2019faceforensics++} rely on hybrid synthesis methods, including traditional graphics (e.g., Face2Face) and learning-based face-swapping. These approaches typically manipulate only facial regions while preserving the original video background\cite{Li_2020_CVPR}, resulting in artifacts such as poor face alignment that simplify detection.

In contrast, our study targets \textbf{fully} generated videos--produced end-to-end by modern video generation models--\textbf{without handcrafted blending or focus on a single object}. These models generate diverse content from general video-caption datasets, making detection more challenging and representative of current generative capabilities.



\subsection{Video Diffusion Models}

Earlier video diffusion models primarily operated in the pixel space~\cite{singer2022makeavideo, ho2022imagen, ho2022video}. Due to computational demands and limited training data, these models struggled to produce high-resolution and coherent videos. With increasing demands for higher resolution and runtime efficiency, latent diffusion models have emerged~\cite{he2023latent,chen2024videocrafter2overcomingdatalimitations,blattmann2023stable,zhang2023i2vgenxl,wang2023lavie,chen2023seine, zhou2023magicvideo,blattmann2023align}. These models substantially reduce computational costs and often adopt a framework similar to Stable Diffusion~\cite{rombach2022highresolution}, enhanced with temporal layers. Based on user input, video generation tasks can be categorized as text-to-video or image-to-video.

\section{Conclusion} \label{sec: conclusion}

Our work is mainly targeting misuse problems in video generation models. We begin by defining the roles present in the real-world setting and, subsequently, design three methods to address misuse issues. Both the detection, source tracing, and prevention tasks utilize the anomalies of spatial-temporal dynamics within the fake videos. Our proposed methods constitute a comprehensive defense pipeline, effectively countering current state-of-the-art video generation models.  

There are some limitations:
The \textit{detection} models and \textit{source tracing} models achieve high accuracy by leveraging features attributed to spatial and temporal spaces. However, the evolution of video generation models (i.e., Sora) will enable the production of more time-consistent and reasonable videos. Our methods may require refinement to detect and trace sources of such advanced fake videos. While the defensive strategies we propose offer effective protection, \textit{directed defense} needs the selection of an appropriate target image for guidance. Conversely, the \textit{undirected defense} may require a larger $\eta$ value for similar defensive effects without needing guidance.
Finally, exploring the abuse concern of the video {\it modification models} is a different task and is not described in our paper.

{\footnotesize \bibliographystyle{acm}
\bibliography{sample}}

\begin{thebibliography}{10}

\bibitem{blattmann2023stable}
{\sc Blattmann, A., Dockhorn, T., Kulal, S., Mendelevitch, D., Kilian, M., Lorenz, D., Levi, Y., English, Z., Voleti, V., Letts, A., Jampani, V., and Rombach, R.}
\newblock Stable video diffusion: Scaling latent video diffusion models to large datasets, 2023.

\bibitem{blattmann2023align}
{\sc Blattmann, A., Rombach, R., Ling, H., Dockhorn, T., Kim, S.~W., Fidler, S., and Kreis, K.}
\newblock Align your latents: High-resolution video synthesis with latent diffusion models, 2023.

\bibitem{bonettini2020video}
{\sc Bonettini, N., Cannas, E.~D., Mandelli, S., Bondi, L., Bestagini, P., and Tubaro, S.}
\newblock Video face manipulation detection through ensemble of cnns, 2020.

\bibitem{carlini2017towards}
{\sc Carlini, N., and Wagner, D.}
\newblock Towards evaluating the robustness of neural networks.
\newblock In {\em 2017 ieee symposium on security and privacy (sp)\/} (2017), Ieee, pp.~39--57.

\bibitem{carreira2018quo}
{\sc Carreira, J., and Zisserman, A.}
\newblock Quo vadis, action recognition? a new model and the kinetics dataset, 2018.

\bibitem{ceylan2023pix2video}
{\sc Ceylan, D., Huang, C.-H.~P., and Mitra, N.~J.}
\newblock Pix2video: Video editing using image diffusion, 2023.

\bibitem{chefer2021transformer}
{\sc Chefer, H., Gur, S., and Wolf, L.}
\newblock Transformer interpretability beyond attention visualization.
\newblock In {\em Proceedings of the IEEE/CVF conference on computer vision and pattern recognition\/} (2021), pp.~782--791.

\bibitem{chen2024videocrafter2overcomingdatalimitations}
{\sc Chen, H., Zhang, Y., Cun, X., Xia, M., Wang, X., Weng, C., and Shan, Y.}
\newblock Videocrafter2: Overcoming data limitations for high-quality video diffusion models, 2024.

\bibitem{chen2023seine}
{\sc Chen, X., Wang, Y., Zhang, L., Zhuang, S., Ma, X., Yu, J., Wang, Y., Lin, D., Qiao, Y., and Liu, Z.}
\newblock Seine: Short-to-long video diffusion model for generative transition and prediction, 2023.

\bibitem{ciftci2020fakecatcher}
{\sc Ciftci, U.~A., Demir, I., and Yin, L.}
\newblock Fakecatcher: Detection of synthetic portrait videos using biological signals.
\newblock {\em IEEE transactions on pattern analysis and machine intelligence\/} (2020).

\bibitem{cozzolino2023synthetic}
{\sc Cozzolino, D., Nagano, K., Thomaz, L., Majumdar, A., and Verdoliva, L.}
\newblock Synthetic image detection: Highlights from the ieee video and image processing cup 2022 student competition, 2023.

\bibitem{dang2020detection}
{\sc Dang, H., Liu, F., Stehouwer, J., Liu, X., and Jain, A.}
\newblock On the detection of digital face manipulation, 2020.

\bibitem{esser2023structure}
{\sc Esser, P., Chiu, J., Atighehchian, P., Granskog, J., and Germanidis, A.}
\newblock Structure and content-guided video synthesis with diffusion models, 2023.

\bibitem{girish2021discovery}
{\sc Girish, S., Suri, S., Rambhatla, S., and Shrivastava, A.}
\newblock Towards discovery and attribution of open-world gan generated images, 2021.

\bibitem{goodfellow2014generative}
{\sc Goodfellow, I.~J., Pouget-Abadie, J., Mirza, M., Xu, B., Warde-Farley, D., Ozair, S., Courville, A., and Bengio, Y.}
\newblock Generative adversarial networks, 2014.

\bibitem{goodfellow2015explaining}
{\sc Goodfellow, I.~J., Shlens, J., and Szegedy, C.}
\newblock Explaining and harnessing adversarial examples, 2015.

\bibitem{gu2021spatiotemporal}
{\sc Gu, Z., Chen, Y., Yao, T., Ding, S., Li, J., Huang, F., and Ma, L.}
\newblock Spatiotemporal inconsistency learning for deepfake video detection.
\newblock In {\em Proceedings of the 29th ACM international conference on multimedia\/} (2021), pp.~3473--3481.

\bibitem{guera2018deepfake}
{\sc G{\"u}era, D., and Delp, E.~J.}
\newblock Deepfake video detection using recurrent neural networks.
\newblock In {\em 2018 15th IEEE international conference on advanced video and signal based surveillance (AVSS)\/} (2018), IEEE, pp.~1--6.

\bibitem{ha2024organic}
{\sc Ha, A. Y.~J., Passananti, J., Bhaskar, R., Shan, S., Southen, R., Zheng, H., and Zhao, B.~Y.}
\newblock Organic or diffused: Can we distinguish human art from ai-generated images?, 2024.

\bibitem{he2021masked}
{\sc He, K., Chen, X., Xie, S., Li, Y., Dollár, P., and Girshick, R.}
\newblock Masked autoencoders are scalable vision learners, 2021.

\bibitem{he2023latent}
{\sc He, Y., Yang, T., Zhang, Y., Shan, Y., and Chen, Q.}
\newblock Latent video diffusion models for high-fidelity long video generation, 2023.

\bibitem{ho2022imagen}
{\sc Ho, J., Chan, W., Saharia, C., Whang, J., Gao, R., Gritsenko, A., Kingma, D.~P., Poole, B., Norouzi, M., Fleet, D.~J., and Salimans, T.}
\newblock Imagen video: High definition video generation with diffusion models, 2022.

\bibitem{ho2020denoising}
{\sc Ho, J., Jain, A., and Abbeel, P.}
\newblock Denoising diffusion probabilistic models, 2020.

\bibitem{ho2022video}
{\sc Ho, J., Salimans, T., Gritsenko, A., Chan, W., Norouzi, M., and Fleet, D.~J.}
\newblock Video diffusion models, 2022.

\bibitem{hu2021dynamic}
{\sc Hu, Z., Xie, H., Wang, Y., Li, J., Wang, Z., and Zhang, Y.}
\newblock Dynamic inconsistency-aware deepfake video detection.
\newblock In {\em IJCAI\/} (2021), pp.~736--742.

\bibitem{hönig2025adversarialperturbationsreliablyprotect}
{\sc Hönig, R., Rando, J., Carlini, N., and Tramèr, F.}
\newblock Adversarial perturbations cannot reliably protect artists from generative ai, 2025.

\bibitem{ji20123d}
{\sc Ji, S., Xu, W., Yang, M., and Yu, K.}
\newblock 3d convolutional neural networks for human action recognition.
\newblock {\em IEEE transactions on pattern analysis and machine intelligence 35}, 1 (2012), 221--231.

\bibitem{karim2023save}
{\sc Karim, N., Khalid, U., Joneidi, M., Chen, C., and Rahnavard, N.}
\newblock Save: Spectral-shift-aware adaptation of image diffusion models for text-driven video editing, 2023.

\bibitem{khan2021video}
{\sc Khan, S.~A., and Dai, H.}
\newblock Video transformer for deepfake detection with incremental learning.
\newblock In {\em Proceedings of the 29th ACM international conference on multimedia\/} (2021), pp.~1821--1828.

\bibitem{kong2024hunyuanvideo}
{\sc Kong, W., Tian, Q., Zhang, Z., Min, R., Dai, Z., Zhou, J., Xiong, J., Li, X., Wu, B., Zhang, J., et~al.}
\newblock Hunyuanvideo: A systematic framework for large video generative models.
\newblock {\em arXiv preprint arXiv:2412.03603\/} (2024).

\bibitem{krishna2023paraphrasing}
{\sc Krishna, K., Song, Y., Karpinska, M., Wieting, J., and Iyyer, M.}
\newblock Paraphrasing evades detectors of ai-generated text, but retrieval is an effective defense.
\newblock {\em arXiv preprint arXiv:2303.13408\/} (2023).

\bibitem{kurakin2017adversarial}
{\sc Kurakin, A., Goodfellow, I., and Bengio, S.}
\newblock Adversarial examples in the physical world, 2017.

\bibitem{Li_2020_CVPR}
{\sc Li, L., Bao, J., Yang, H., Chen, D., and Wen, F.}
\newblock Advancing high fidelity identity swapping for forgery detection.
\newblock In {\em Proceedings of the IEEE/CVF Conference on Computer Vision and Pattern Recognition (CVPR)\/} (June 2020).

\bibitem{li2021deepfake}
{\sc Li, M., Liu, B., Hu, Y., Zhang, L., and Wang, S.}
\newblock Deepfake detection using robust spatial and temporal features from facial landmarks.
\newblock In {\em 2021 IEEE International Workshop on Biometrics and Forensics (IWBF)\/} (2021), IEEE, pp.~1--6.

\bibitem{liang2023adversarial}
{\sc Liang, C., Wu, X., Hua, Y., Zhang, J., Xue, Y., Song, T., Xue, Z., Ma, R., and Guan, H.}
\newblock Adversarial example does good: preventing painting imitation from diffusion models via adversarial examples.
\newblock In {\em Proceedings of the 40th International Conference on Machine Learning\/} (2023), pp.~20763--20786.

\bibitem{liu2023videop2p}
{\sc Liu, S., Zhang, Y., Li, W., Lin, Z., and Jia, J.}
\newblock Video-p2p: Video editing with cross-attention control, 2023.

\bibitem{lu2023seeing}
{\sc Lu, Z., Huang, D., Bai, L., Qu, J., Wu, C., Liu, X., and Ouyang, W.}
\newblock Seeing is not always believing: Benchmarking human and model perception of ai-generated images, 2023.

\bibitem{ma2025step}
{\sc Ma, G., Huang, H., Yan, K., Chen, L., Duan, N., Yin, S., Wan, C., Ming, R., Song, X., Chen, X., et~al.}
\newblock Step-video-t2v technical report: The practice, challenges, and future of video foundation model.
\newblock {\em arXiv preprint arXiv:2502.10248\/} (2025).

\bibitem{ma2022xclip}
{\sc Ma, Y., Xu, G., Sun, X., Yan, M., Zhang, J., and Ji, R.}
\newblock X-clip: End-to-end multi-grained contrastive learning for video-text retrieval, 2022.

\bibitem{madry2019deep}
{\sc Madry, A., Makelov, A., Schmidt, L., Tsipras, D., and Vladu, A.}
\newblock Towards deep learning models resistant to adversarial attacks, 2019.

\bibitem{mitchell2023detectgpt}
{\sc Mitchell, E., Lee, Y., Khazatsky, A., Manning, C.~D., and Finn, C.}
\newblock Detectgpt: Zero-shot machine-generated text detection using probability curvature, 2023.

\bibitem{molad2023dreamix}
{\sc Molad, E., Horwitz, E., Valevski, D., Acha, A.~R., Matias, Y., Pritch, Y., Leviathan, Y., and Hoshen, Y.}
\newblock Dreamix: Video diffusion models are general video editors, 2023.

\bibitem{moosavidezfooli2016deepfool}
{\sc Moosavi-Dezfooli, S.-M., Fawzi, A., and Frossard, P.}
\newblock Deepfool: a simple and accurate method to fool deep neural networks, 2016.

\bibitem{Mullan-Hotshot-XL-2023}
{\sc Mullan, J., Crawbuck, D., and Sastry, A.}
\newblock {Hotshot-XL}, Oct. 2023.

\bibitem{nan2025openvid1mlargescalehighqualitydataset}
{\sc Nan, K., Xie, R., Zhou, P., Fan, T., Yang, Z., Chen, Z., Li, X., Yang, J., and Tai, Y.}
\newblock Openvid-1m: A large-scale high-quality dataset for text-to-video generation, 2025.

\bibitem{ni2022core}
{\sc Ni, Y., Meng, D., Yu, C., Quan, C., Ren, D., and Zhao, Y.}
\newblock Core: Consistent representation learning for face forgery detection, 2022.

\bibitem{papernot2015limitations}
{\sc Papernot, N., McDaniel, P., Jha, S., Fredrikson, M., Celik, Z.~B., and Swami, A.}
\newblock The limitations of deep learning in adversarial settings, 2015.

\bibitem{radford2021learning}
{\sc Radford, A., Kim, J.~W., Hallacy, C., Ramesh, A., Goh, G., Agarwal, S., Sastry, G., Askell, A., Mishkin, P., Clark, J., Krueger, G., and Sutskever, I.}
\newblock Learning transferable visual models from natural language supervision, 2021.

\bibitem{rombach2022highresolution}
{\sc Rombach, R., Blattmann, A., Lorenz, D., Esser, P., and Ommer, B.}
\newblock High-resolution image synthesis with latent diffusion models, 2022.

\bibitem{rossler2019faceforensics++}
{\sc Rossler, A., Cozzolino, D., Verdoliva, L., Riess, C., Thies, J., and Nie{\ss}ner, M.}
\newblock Faceforensics++: Learning to detect manipulated facial images.
\newblock In {\em Proceedings of the IEEE/CVF international conference on computer vision\/} (2019), pp.~1--11.

\bibitem{Selvaraju_2019}
{\sc Selvaraju, R.~R., Cogswell, M., Das, A., Vedantam, R., Parikh, D., and Batra, D.}
\newblock Grad-cam: Visual explanations from deep networks via gradient-based localization.
\newblock {\em International Journal of Computer Vision 128}, 2 (Oct. 2019), 336–359.

\bibitem{sha2023defake}
{\sc Sha, Z., Li, Z., Yu, N., and Zhang, Y.}
\newblock De-fake: Detection and attribution of fake images generated by text-to-image generation models, 2023.

\bibitem{shan2023glaze}
{\sc Shan, S., Cryan, J., Wenger, E., Zheng, H., Hanocka, R., and Zhao, B.~Y.}
\newblock Glaze: Protecting artists from style mimicry by text-to-image models, 2023.

\bibitem{shin2023editavideo}
{\sc Shin, C., Kim, H., Lee, C.~H., gil Lee, S., and Yoon, S.}
\newblock Edit-a-video: Single video editing with object-aware consistency, 2023.

\bibitem{singer2022makeavideo}
{\sc Singer, U., Polyak, A., Hayes, T., Yin, X., An, J., Zhang, S., Hu, Q., Yang, H., Ashual, O., Gafni, O., Parikh, D., Gupta, S., and Taigman, Y.}
\newblock Make-a-video: Text-to-video generation without text-video data, 2022.

\bibitem{szegedy2014intriguing}
{\sc Szegedy, C., Zaremba, W., Sutskever, I., Bruna, J., Erhan, D., Goodfellow, I., and Fergus, R.}
\newblock Intriguing properties of neural networks, 2014.

\bibitem{thies2016face2face}
{\sc Thies, J., Zollhofer, M., Stamminger, M., Theobalt, C., and Nie{\ss}ner, M.}
\newblock Face2face: Real-time face capture and reenactment of rgb videos.
\newblock In {\em Proceedings of the IEEE conference on computer vision and pattern recognition\/} (2016), pp.~2387--2395.

\bibitem{tong2022videomae}
{\sc Tong, Z., Song, Y., Wang, J., and Wang, L.}
\newblock Videomae: Masked autoencoders are data-efficient learners for self-supervised video pre-training, 2022.

\bibitem{tramèr2017space}
{\sc Tramèr, F., Papernot, N., Goodfellow, I., Boneh, D., and McDaniel, P.}
\newblock The space of transferable adversarial examples, 2017.

\bibitem{wang2020fakespotter}
{\sc Wang, R., Juefei-Xu, F., Ma, L., Xie, X., Huang, Y., Wang, J., and Liu, Y.}
\newblock Fakespotter: A simple yet robust baseline for spotting ai-synthesized fake faces, 2020.

\bibitem{wang2020cnngenerated}
{\sc Wang, S.-Y., Wang, O., Zhang, R., Owens, A., and Efros, A.~A.}
\newblock Cnn-generated images are surprisingly easy to spot... for now, 2020.

\bibitem{wang2023videofactory}
{\sc Wang, W., Yang, H., Tuo, Z., He, H., Zhu, J., Fu, J., and Liu, J.}
\newblock Videofactory: Swap attention in spatiotemporal diffusions for text-to-video generation, 2023.

\bibitem{wang2023videocomposer}
{\sc Wang, X., Yuan, H., Zhang, S., Chen, D., Wang, J., Zhang, Y., Shen, Y., Zhao, D., and Zhou, J.}
\newblock Videocomposer: Compositional video synthesis with motion controllability, 2023.

\bibitem{wang2023lavie}
{\sc Wang, Y., Chen, X., Ma, X., Zhou, S., Huang, Z., Wang, Y., Yang, C., He, Y., Yu, J., Yang, P., Guo, Y., Wu, T., Si, C., Jiang, Y., Chen, C., Loy, C.~C., Dai, B., Lin, D., Qiao, Y., and Liu, Z.}
\newblock Lavie: High-quality video generation with cascaded latent diffusion models, 2023.

\bibitem{wang2024internvid}
{\sc Wang, Y., He, Y., Li, Y., Li, K., Yu, J., Ma, X., Li, X., Chen, G., Chen, X., Wang, Y., He, C., Luo, P., Liu, Z., Wang, Y., Wang, L., and Qiao, Y.}
\newblock Internvid: A large-scale video-text dataset for multimodal understanding and generation, 2024.

\bibitem{wodajo2023deepfake}
{\sc Wodajo, D., Atnafu, S., and Akhtar, Z.}
\newblock Deepfake video detection using generative convolutional vision transformer, 2023.

\bibitem{wu2023tuneavideo}
{\sc Wu, J.~Z., Ge, Y., Wang, X., Lei, W., Gu, Y., Shi, Y., Hsu, W., Shan, Y., Qie, X., and Shou, M.~Z.}
\newblock Tune-a-video: One-shot tuning of image diffusion models for text-to-video generation, 2023.

\bibitem{xiao2024videodiffusionmodelstrainingfree}
{\sc Xiao, Z., Zhou, Y., Yang, S., and Pan, X.}
\newblock Video diffusion models are training-free motion interpreter and controller, 2024.

\bibitem{yu2019attributing}
{\sc Yu, N., Davis, L., and Fritz, M.}
\newblock Attributing fake images to gans: Learning and analyzing gan fingerprints, 2019.

\bibitem{zhang2023show1}
{\sc Zhang, D.~J., Wu, J.~Z., Liu, J.-W., Zhao, R., Ran, L., Gu, Y., Gao, D., and Shou, M.~Z.}
\newblock Show-1: Marrying pixel and latent diffusion models for text-to-video generation, 2023.

\bibitem{zhang2023i2vgenxl}
{\sc Zhang, S., Wang, J., Zhang, Y., Zhao, K., Yuan, H., Qin, Z., Wang, X., Zhao, D., and Zhou, J.}
\newblock I2vgen-xl: High-quality image-to-video synthesis via cascaded diffusion models, 2023.

\bibitem{zhang2023controlvideo}
{\sc Zhang, Y., Wei, Y., Jiang, D., Zhang, X., Zuo, W., and Tian, Q.}
\newblock Controlvideo: Training-free controllable text-to-video generation, 2023.

\bibitem{zhao2023controlvideo}
{\sc Zhao, M., Wang, R., Bao, F., Li, C., and Zhu, J.}
\newblock Controlvideo: Conditional control for one-shot text-driven video editing and beyond, 2023.

\bibitem{zhao2023motiondirector}
{\sc Zhao, R., Gu, Y., Wu, J.~Z., Zhang, D.~J., Liu, J., Wu, W., Keppo, J., and Shou, M.~Z.}
\newblock Motiondirector: Motion customization of text-to-video diffusion models, 2023.

\bibitem{zhou2023magicvideo}
{\sc Zhou, D., Wang, W., Yan, H., Lv, W., Zhu, Y., and Feng, J.}
\newblock Magicvideo: Efficient video generation with latent diffusion models, 2023.

\end{thebibliography}

\appendices

\section{More Details for Diffusion Model} \label{appendix:diffusion}

In this section, we want to discuss more details about the diffusion model. Since we have already explained what the diffusion process is in~\autoref{sec:background}. Therefore, In this part, we will mainly focus on the reverse (denoising) process.
The reverse process can be described as:
\begin{align*}
    p_\theta(x_{0:T}) = p(x_{T})\prod_{t=1}^{T} p_\theta(x_{t-1}|x_t)
\end{align*} 
where $x_{T} \sim \mathcal{N} (0, I)$ and $x_0$ is the denoised image. For a step $t \in [0,T]$, the noise image $x_{t-1}$ denoising from $x_t$ can be represented as:  
\begin{align*}
    p_\theta(x_{t-1}|x_t) = \mathcal{N}(x_{t-1};{\boldsymbol{\mu}}_\theta(x_t,t),{\boldsymbol{\Sigma}}_\theta(x_t,t))
\end{align*}

The ground truth denoised image $x_{t-1}$ can be sample from distribution $\mathcal{N}(x_{t-1};{\boldsymbol{\bar{\mu}}}(x_t,x_0),\boldsymbol{\bar{\beta}_t} \mathbf{I})$. In DDPM~\cite{ho2020denoising}, $\boldsymbol{\Sigma}_\theta(x_t,t)$ is set to $\sigma^2_t \mathbf{I}$ and is untrainable. Therefore, the diffusion model is mainly to approximate $\boldsymbol{\bar{\mu}}(x_t,x_0)$ using ${\boldsymbol{\mu}}_\theta(x_t,t)$. After applying Bayes's rule to expend $\boldsymbol{\bar{\mu}}(x_t,x_0)$. we can get

\begin{align}
   &\boldsymbol{\bar{\mu}}_t = \frac{\sqrt{\alpha_t}(1-\bar{\alpha}_{t-1})}{1-\bar{\alpha}_t}x_t + \nonumber \\
   &\frac{\sqrt{\bar{\alpha}_{t-1}}(1-\alpha_t)}{1-\bar{\alpha}_t}\frac{1}{\sqrt{\bar{\alpha}_t}}(x_t-\sqrt{1-\bar{\alpha}_t}\epsilon_t) \label{eq:mu_t}
\end{align}

Because we already have the ground-truth $\boldsymbol{\bar{\mu}}_t$ the initial objective function can be written as:
\begin{equation}
    L_t(\theta) = \mathbb{E}_{x_0,\epsilon_t} \left[ \lVert \boldsymbol{\bar{\mu}}_t - \boldsymbol{\mu}_\theta(\sqrt{\bar{\alpha}_t} x_0 + \sqrt{1 - \bar{\alpha}_t} \epsilon_t, t) \rVert^2_2\right]\;
    \label{eq:initial_eq}
\end{equation}

In~\autoref{eq:mu_t}, we applied~\autoref{x_t} to represent $x_0$ in $\boldsymbol{\bar{\mu}}(x_t,x_0)$. The only term that is unknown and predictable is $\epsilon_t$. Thus, ${\boldsymbol{\mu}}_\theta(x_t,t)$ is reform as:
\begin{align*}
   &\boldsymbol{\mu}_{\theta}(x_t,t) = \frac{\sqrt{\alpha_t}(1-\bar{\alpha}_{t-1})}{1-\bar{\alpha}_t}x_t + \\ \nonumber &\frac{\sqrt{\bar{\alpha}_{t-1}}(1-\alpha_t)}{1-\bar{\alpha}_t}\frac{1}{\sqrt{\bar{\alpha}_t}}(x_t-\sqrt{1-\bar{\alpha}_t}\epsilon_t(x_t,t)) 
\end{align*}

The training object of predict $\boldsymbol{\mu}_{\theta}(x_t,t)$ approximate $\boldsymbol{\bar{\mu}}(x_t,x_0)$ can then replace by predict $\epsilon_t$ given $x_t$ and $t$. Finally, after disregarding certain coefficient terms, we obtain the loss function in the form of~\autoref{eq:loss}, derived from our initial objective function presented in~\autoref{eq:initial_eq}.

\section{More Details for Fake Video Detection}

We showed more experiments results for our detection models.

\begin{table}[!h]
    \centering
    \belowrulesep=0pt
    \aboverulesep=0pt
    \caption{FPR/FNR for \textit{detection} on OpenVid-1M}
    \label{tab:appendix_fake_video_openvid}
    \resizebox{0.47\textwidth}{!}{
    \centering
    \begin{tabular}{c|ccc}
    \toprule
         \multirow{2}{*}{Model} & \multicolumn{3}{c}{OpenVid-1M}\\ \cmidrule(lr){2-4} 
         & \id & \xclip & \mae \\ \midrule 
         Hunyuan& $0.05/0.12$ & $0.08/0.01$ & $0.04/0.01$ \\
         VGen(T2V)& $0.06/0.04$ & $0.01/0.01$ & $0.01/0.01$ \\
         VGen(I2V)& $0.07/0.25$ & $0.05/0.07$ &  $0.02/0.01$\\
         LaVie& $0.07/0.15$ & $0.01/0.01$ & $0.03/0.02$ \\
         Seine& $0.14/0.35$ & $0.17/0.29$ & $0.01/0.01$ \\
         StepVideo& $0.04/0.05$ & $0.12/0.15$ & $0.01/0.01$ \\
         SVD& $0.11/0.14$ & $0.05/0.04$ & $0.01/0.01$ \\
         VideoCrafter(T2V)& $0.07/0.03$ & $0.06/0.01$ & $0.01/0.01$ \\
         VideoCrafter(I2V)&$0.20/0.10$ & $0.01/0.02$ & $0.01/0.01$ \\
    \bottomrule
    \end{tabular}}
\end{table}

\begin{table}[!h]
    \centering
    \belowrulesep=0pt
    \aboverulesep=0pt
    \caption{FPR/FNR for \textit{detection} on InternVid.}
    \label{tab:appendix_fake_video_internvid}
    \resizebox{0.47\textwidth}{!}{
    \centering
    \begin{tabular}{c|ccc}
    \toprule
         \multirow{2}{*}{Model} & \multicolumn{3}{c}{InternVid}\\ \cmidrule(lr){2-4}
         & \id & \xclip & \mae  \\ \midrule 
         Hunyuan& $0.10/0.07$ & $0.02/0.02$ & $0.02/0.01$ \\
         VGen (T2V)& $0.07/0.10$ & $0.25/0.33$ & $0.01/0.01$  \\
         VGen (I2V)& $0.06/0.31$ & $0.16/0.29$ & $0.01/0.02$ \\
         LaVie& $0.08/0.24$ & $0.16/0.44$ & $0.04/0.01$  \\
         Seine& $0.04/0.24$ & $0.11/0.19$ & $0.01/0.01$ \\
         StepVideo& $0.05/0.11$ & $0.07/0.12$ & $0.03/0.03$ \\
         SVD& $0.17/0.15$ & $0.10/0.55$ & $0.08/0.02$ \\
         VideoCrafter (T2V)& $0.11/0.10$ & $0.08/0.19$ & $0.01/0.01$\\
         VideoCrafter (I2V)& $0.15/0.23$ & $0.24/0.38$ & $0.01/0.01$\\
    \bottomrule
    \end{tabular}}
\end{table}

\section{More Details for Misuse Prevention \ding{182}}

\begin{table}[!h]
    \centering
    \caption{The defensive effectiveness of adversarial examples with four varying levels of perturbation intensity was evaluated on Stable Video Diffusion, Gen-2, and Pika Lab. \ding{51}: motion prediction is reasonable \ding{55}: motion prediction is distorted.}
    \label{tab:results_defense}
    \resizebox{0.47\textwidth}{!}{
    \Large
    \begin{tabular}{ccccccccccccc}
    \toprule[1.5pt]
       \multirow{2}{*}{\begin{tabular}[c]{@{}c@{}} {} \end{tabular}}  &\multicolumn{4}{c}{SVD} & \multicolumn{4}{c}{Gen-2}& \multicolumn{4}{c}{Pika Lab} \\
       \cmidrule(lr){2-5}\cmidrule(lr){6-9} \cmidrule(lr){10-13} 
         & $\frac{2}{255}$ & $\frac{4}{255}$ & $\frac{8}{255}$ & $\frac{16}{255}$ & $\frac{2}{255}$ & $\frac{4}{255}$ & $\frac{8}{255}$ & $\frac{16}{255}$& $\frac{2}{255}$ & $\frac{4}{255}$ & $\frac{8}{255}$ &$\frac{16}{255}$ \\
         \midrule
          Baseline & \ding{55} & \ding{55} & \ding{51} & \ding{51} & \ding{55} & \ding{55} & \ding{51} & \ding{51} & \ding{55} & \ding{55} & \ding{55} & \ding{51} \\
          \rowcolor{gray!25} Ours (directed)& \ding{55} & \ding{51} & \ding{51} & \ding{51} & \ding{55} & \ding{51} & \ding{51} & \ding{51} & \ding{55} & \ding{55} & \ding{51} & \ding{51} \\
          \rowcolor{gray!25} Ours (undirected)& \ding{55} & \ding{51} & \ding{51} & \ding{51} & \ding{55} & \ding{55} & \ding{51} & \ding{51} & \ding{55} & \ding{55} & \ding{55} & \ding{51} \\
    \bottomrule[1.5pt]
    \end{tabular}}
\end{table}

The results of each defense method under four different $\eta$ settings are shown in~\autoref{tab:results_defense}. Both proposed methods effectively prevent Stable Video Diffusion~\cite{blattmann2023stable} from generating regular videos. We used the image and video encoders of Stable Video Diffusion in both \textit{directed defense} and \textit{undirected defense}. When dealing with unknown video generation models, the adversarial examples generated by the \textit{directed defense} method have weaker preventive capabilities.

\newpage
\onecolumn
\section{More Details for Visualization} \label{sec:appendix_detection}

\begin{figure*}[!h]
\centering
    \begin{subfigure}{\textwidth}
    \centering
        \includegraphics[width=0.85\textwidth]{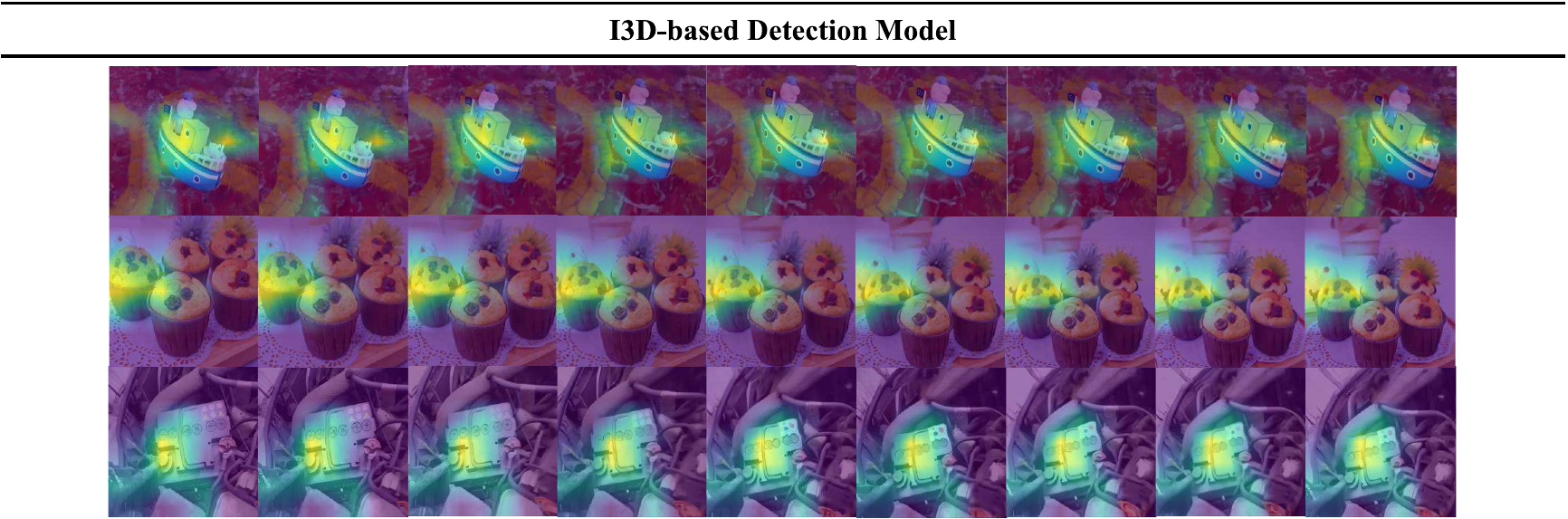}
    \end{subfigure}
    \\
    \vspace{0.1cm}
    \begin{subfigure}{\textwidth}
    \centering
        \includegraphics[width=0.85\textwidth]{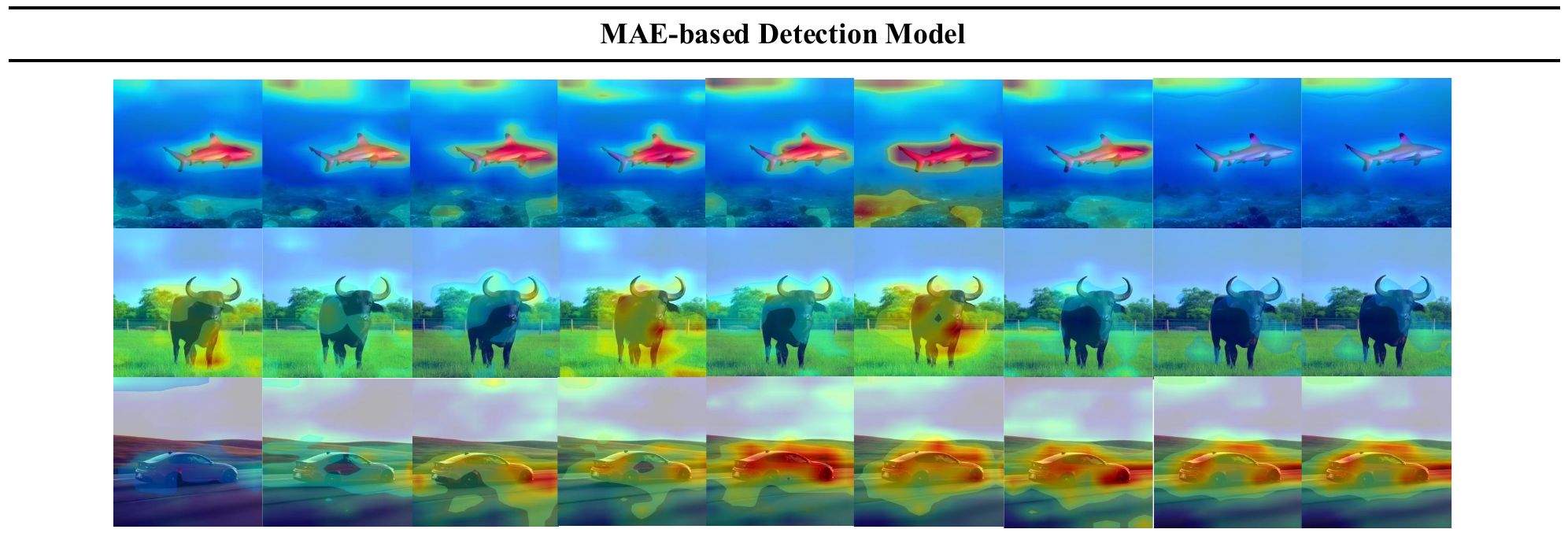}
    \end{subfigure}
    \caption{A complete video featuring heatmaps using Grad-CAM on both \id-based and \mae-based detection models. \mae-based detection model can detect more abnormal points in one video while \id-based detection model can only detect one place.}
    \label{fig:appendix_detection}
\end{figure*}

In~\autoref{fig:appendix_detection}, we provide several video samples with Grad-CAM. By comparing the \id-based detection model with the \mae-based detection model. We found that \mae-based detection models are more agile than \id-based. It can detect inconsistent features in spatial and temporal domains. Therefore achieving a higher detection accuracy. 

\begin{figure*}[!h]
    \centering
    \includegraphics[width=0.85\textwidth]{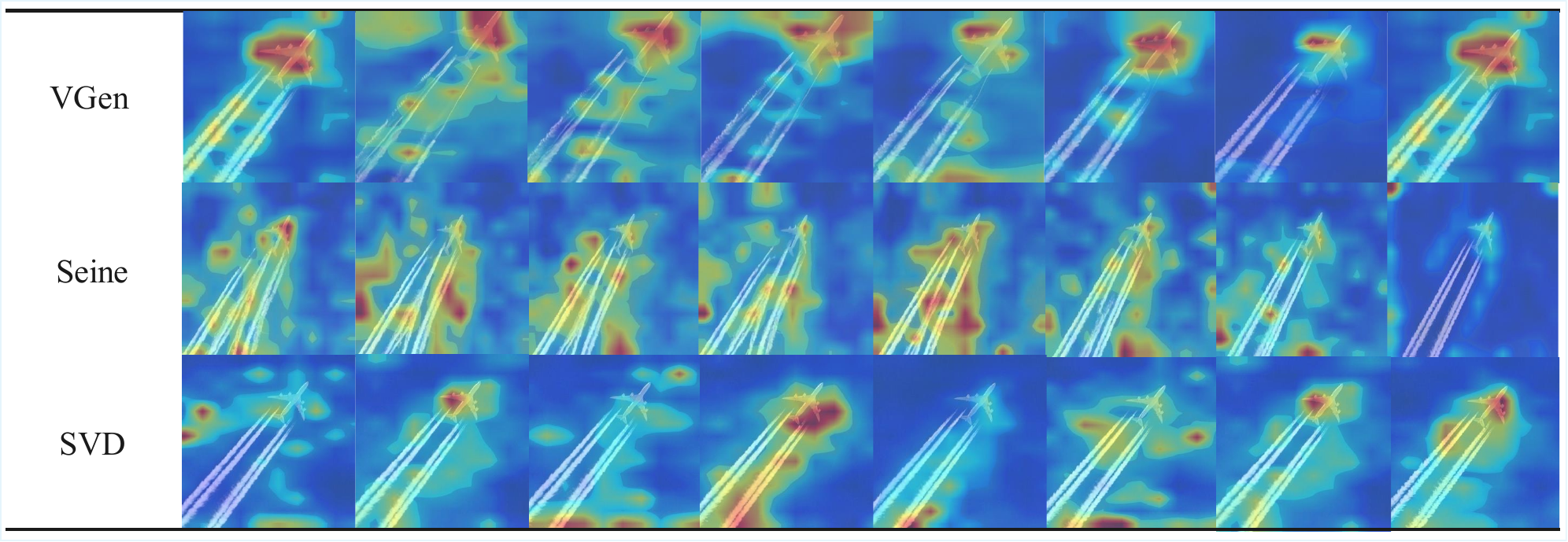}
    \caption{Source tracing model relies on detecting characteristics at different positions in the video to determine the generating model. Note: These models are all image-to-video and do not need prompts.}
    \label{fig:source_tracing_ablation}
\end{figure*}

In~\autoref{fig:source_tracing_ablation}, we visualize the outputs of the \mae-based source tracing model for videos generated by different models. We observe that the attributor still focuses on the primary generated objects to make its predictions.

\newpage
\section{More Details for Misuse Prevention \ding{183}}

\begin{figure*}[h]
    \centering
    \begin{subfigure}{\textwidth}
    \centering
        \includegraphics[width=0.9\textwidth]{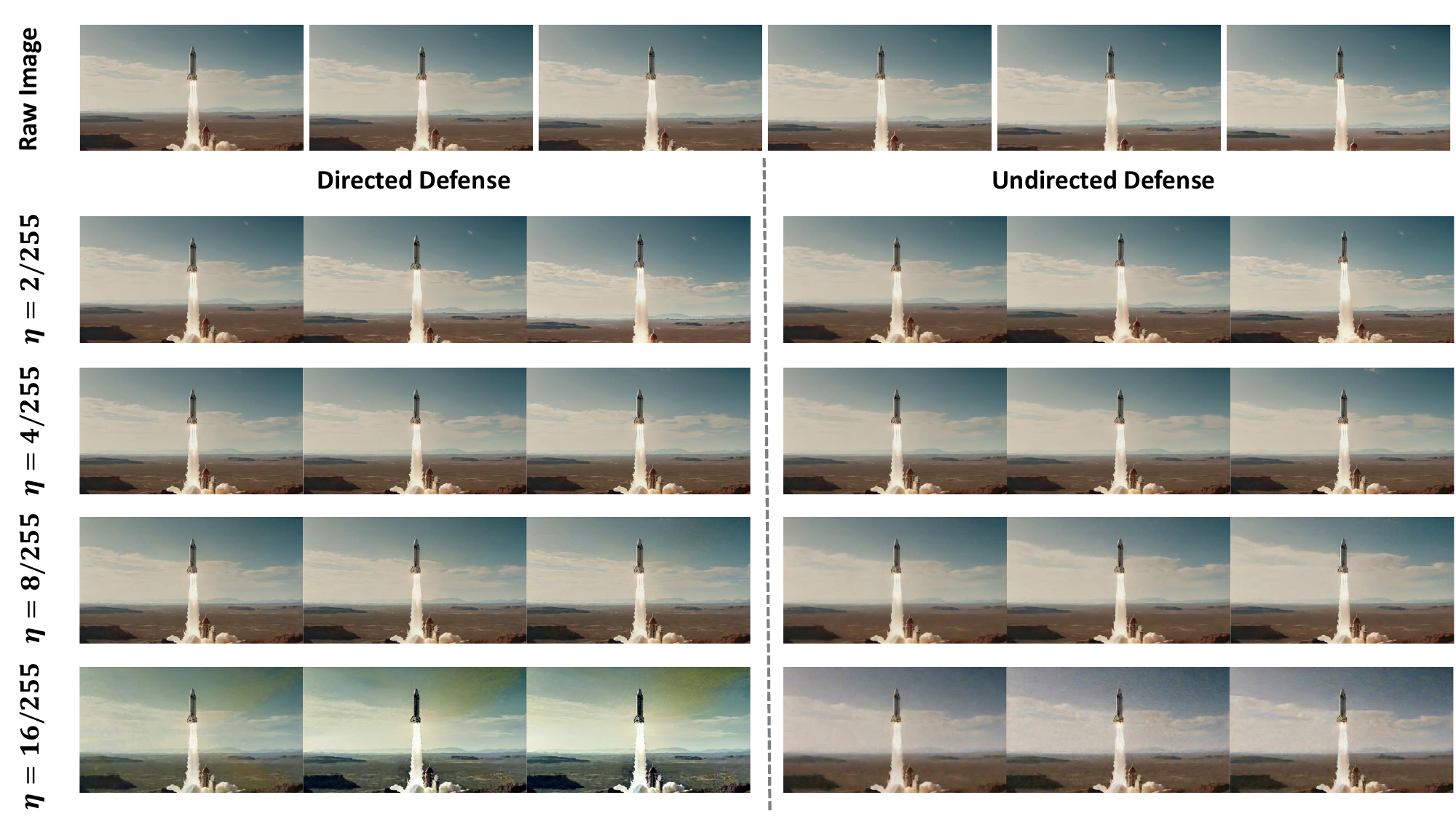}
        \caption{Demonstrate the defensive efficacy of \textit{directed defense} and \textit{undirected defense} under various $\eta$ settings.}
        \label{fig:appendix_shield}
    \end{subfigure}
    \begin{subfigure}{\textwidth}
        \centering
        \includegraphics[width=0.9\textwidth]{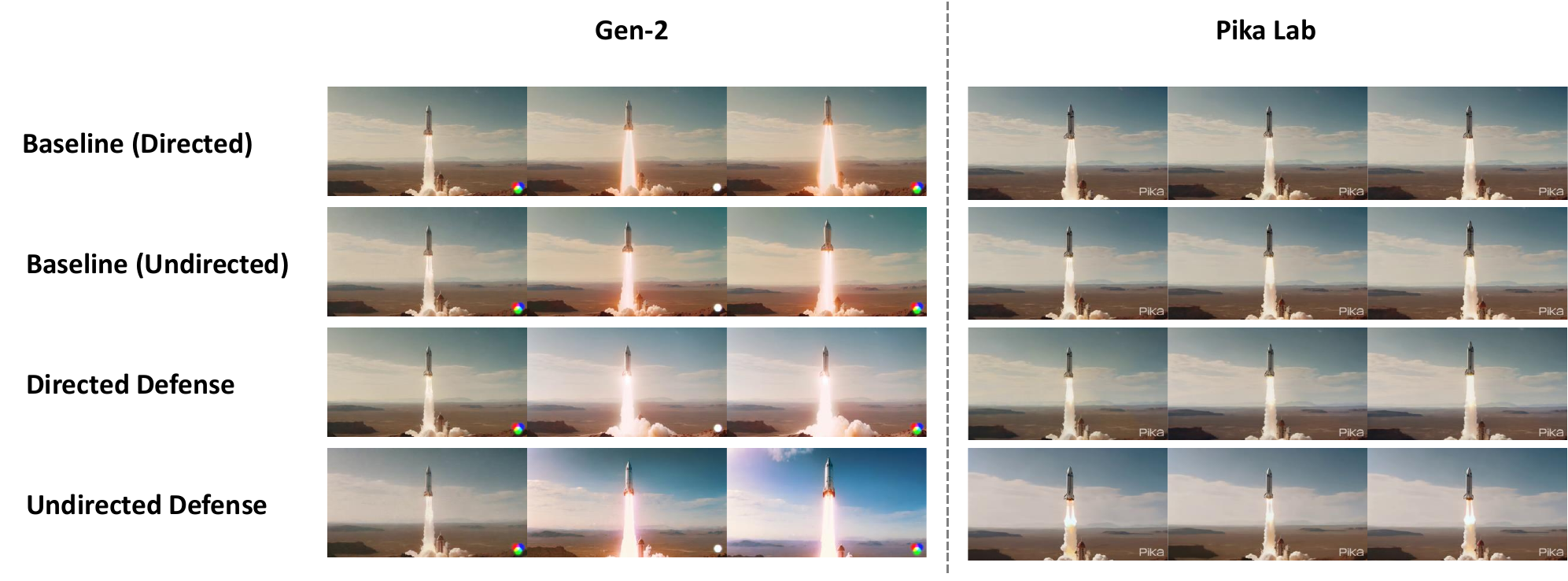}
        \caption{Comparision between our \textit{directed defense} and \textit{undirected defense} with baseline methods.}
        \label{fig:appendix_compare}
    \end{subfigure}
    \caption{Present two defense strategies across various parameters and against different generation models.}
    \label{fig:appendix_misuse}
\end{figure*}

We show the supplementary image in~\autoref{fig:appendix_misuse}.~\autoref{fig:appendix_shield} showcase \textit{adversarial examples} generated from \textit{directed defense} and \textit{undirected defense}. In~\autoref{fig:appendix_compare}, it is evident that for Gen-2, the two baseline methods fail to disrupt the generation process significantly. Notably, the rocket in the video continues to ascend seamlessly into the sky, and the smoke trailing the rocket behaves logically. Our \textit{directed defense} method can effectively immobilize the rocket in mid-air, while the \textit{undirected defense} method. However, it does not significantly interfere with the rocket's movement and can compromise the overall coherence of the video. Conducting tests on the Pika Lab platform revealed that the vanilla videos generated by the model already depict the rocket as stationary. After applying the two baseline methods, the video has no discernible change compared to the original. However, our \textit{directed defense} method succeeds in freezing the motion of the smoke emitted by the rocket, thereby further undermining the video's logical integrity.

\end{document}